\documentclass[preprint,12pt]{elsarticle}

\usepackage{amssymb}
\usepackage{amsmath}

\usepackage{url}

\journal{J Comp Phys}

\begin{document}

\begin{frontmatter}



\title{Mixed Cayley and Cartesian Sampling for Fast and Accurate Coverage and Configurational Entropy Computation}


\author{Yichi Zhang, Meera Sitharam} 

\affiliation{organization={Department of CISE, University of Florida},
            addressline={1889 Museum Rd.}, 
            city={Gainesville},
            postcode={32611}, 
            state={FL},
            country={US}}

\begin{abstract}
This article describes Uniform Cartesian (UC), a   \emph{deterministic} methodology for computing configurational entropy via relative volume of energy basins, for pair-potential assembly systems. The methodology, aimed at navigating the twin curses of  dimensionality and topological complexity of  constrained configurational regions, adapts to longer ranged pair potentials. 

The methodology significantly extends EASAL (Efficient Atlasing and Sampling of Assembly Landscapes), a recent methodology 
based on modern discrete geometry concepts.  Key features include atlasing configurational regions and sampling in a distance-based Cayley coordinate parametrization. 
 This achieves avoidance of gradient descent and retraction maps  used by prevailing methods to sample a nonlinear, constrained configurational regions that are of effectively lower intrinsic dimension,  within high ambient dimension defined by available degrees of freedom.
  Furthermore, UC  iteratively maps between Cayley and Cartesian coordinates, avoiding matrix derivative (Jacobian and Hessian computations, subject to ill-conditioning and numerical errors) customarily used for conversion between internal (Cayley) and Cartesian coordinates.

 We guarantee correctness, optimal time and space complexity, and efficiency-accuracy tradeoffs   using  rigorous algorithmic analysis. 
An opensource software implementation 
is  provided on top of previously curated EASAL opensource software. 
%
Running on one 
CPU node with 40GB of memory, variants of UC  accurately calculate the relative volume of standard energy basins within hours, even without parallelization.  

 This article's emphasis is not extensive benchmark comparisons of  large-scale parallel implementations or prevailing methods. Rather,   proof-of-concept demonstrations of the unique features of  UC are given along with  test-case comparisons   between the mainstream Monte Carlo method, the new UC variants, and  the ``vanilla'' EASAL. 

\end{abstract}

\begin{graphicalabstract}
\includegraphics[width=\textwidth]{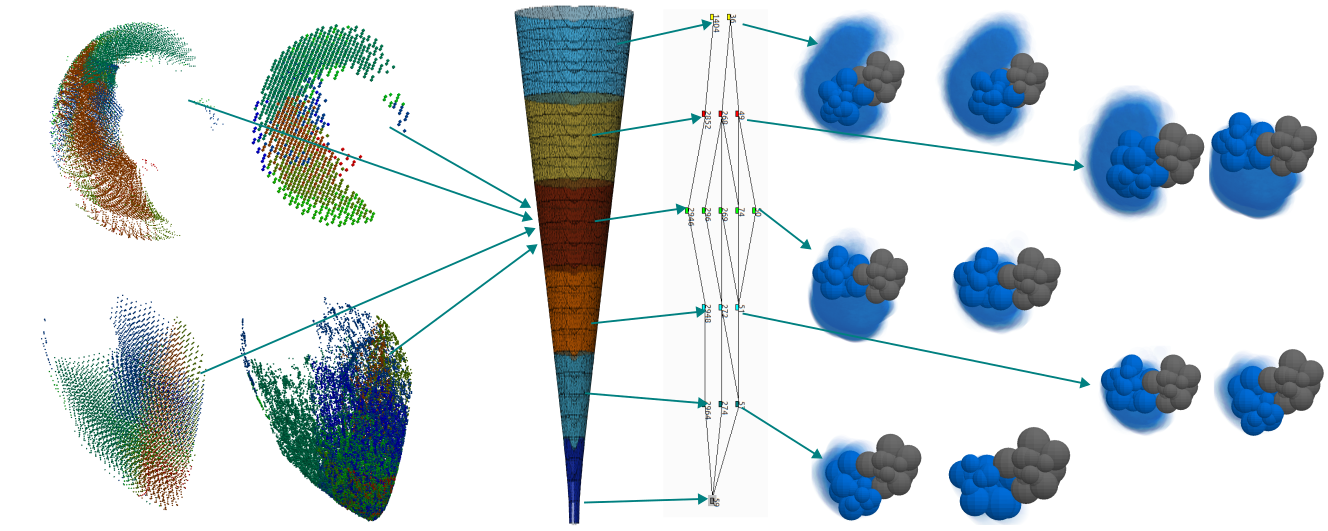}
\end{graphicalabstract}

\begin{highlights}
\item Deterministic discrete geometry method for configurational entropy
\item Mitigating curses of dimensionality and topological complexity of configuration space
\item Iterative maps between Cayley and Cartesian coordinates to avoid numerical errors
\item Optimal time and space complexity, rigorous accuracy and efficiency guarantees  
\item Curated opensource software, favorable performance comparison against Monte Carlo
\end{highlights}


\end{frontmatter}



\section{Introduction}

We present a deterministic, discrete-geometry based methodology UC (for \textit{Uniform Cartesian}) that computes configurational entropy (a discrete, relative volume measure) for a   well-defined and commonly occurring class of pair-potential driven, softmatter assembly landscapes,   with  guaranteed  accuracy, in (sub) linear output complexity.

Access to an accurate free energy landscape is key for molecular science and a vital aspect of free energy calculation is configurational entropy. Specifically, for systems whose free energy landscape consists of multiple energy basins, configurational entropy computations reduce to computing ratios of basin volumes \cite{Casiulis2022}.  Free energy and configurational entropy computation \cite{ZhouGilson2009} of guaranteed accuracy challenge prevailing randomized methods based on Molecular Dynamics or Monte Carlo \cite{Baker1999, WalesBogdan2006, Rybkin2013, BussiBranduardi2015, Li2023} that are not known to be ergodic. These problems persist even for relatively simple systems of soft-matter assembly driven by (short-ranged) Lennard-Jones potentials. For such assembly systems, configurational entropy computation involves two crucial tasks, both of which we treat in detail: (a) accurate topological organization of constant potential energy regions within and between basins based on both the number and size of energy barriers between the regions (b) uniformly sampling and computing discrete (relative) volume measures of low constant energy, effectively lower dimensional and topologically complex configurational regions   within higher dimensional, high energy ambient spaces.

We focus on assembly landscapes - configuration spaces - of systems of multiple internally rigid molecular components. 
Flexible components can be modeled as assembly systems of their maximal rigid components using tethering constraints. We formulate the landscapes  as feasible regions of distance constraint systems between \textit{point sets} by geometrizing the pair-potentials. The assembly landscape   consists of configurations of a finite collection of (internally rigid) point sets in $\mathbb{R}^d$ that satisfy a distance constraint system (equalities and/or inequalities) between point-pairs in different point sets. Each point set is treated as an equivalence class modulo translations and rotations. Formally, a configuration is a collection of orientations (translation/rotation) that relatively position each point set with respect to a fixed point set while satisfying the constraints. Such systems are not limited to molecular or soft-matter science; they can be used in the study of kinematic mechanisms, (under-constrained) mechanical CAD designs, metamaterials, etc. 

The new UC methodology deterministically   computes a discrete volume measure via discrete uniform sampling of a large, well-defined and commonly occurring class of distance constrained configuration spaces with  guaranteed   accuracy in (sub)linear output complexity.
The new methodology significantly extends ``vanilla'' EASAL (efficient atlasing and sampling of assembly landscapes) \cite{SitharamGao2010, Ozkan2018ACMTOMS, Prabhu2020JCIM, Wu2020PLoS} described in Section \ref{sec:previous}, a state-of-the-art discrete geometry-based methodology for roadmapping, sampling and analyzing the landscape of macromolecular configurations satisfying possible sets of pairwise inter-molecular distance constraints \cite{OzkanBiCoB2011, Prabhu2020JCIM}. EASAL is both a resource-light, stand-alone method and also complements prevailing MC, MD and docking methods\cite{Ozkan2021JCTC}, demonstrating superior performance, especially for discontinuous pair-potential energy landscapes. 

The EASAL methodology \cite{Prabhu2020JCIM} and curated open-source software \cite{Ozkan2018ACMTOMS} can efficiently generate an exhaustive ensemble of structures lying within specific pair-potential wells, discretized as a staircase of nested distance-interval constraints.  \\(See \ref{sec:guide}: \url{https://bitbucket.org/geoplexity/easal-dev/src/master/}; UC in branch ``feature/UniformCartesian''; and  \url{http://www.cise.ufl.edu/~sitharam/EASALvideo.mpeg} for video demo of ``vanilla'' EASAL)  The methodology has been earlier used for different biomolecular applications, including effectively predicting crucial interactions for autonomous viral shell assembly \cite{Wu2020PLoS}, sticky-sphere path integrals \cite{Prabhu2020JCIM} and structure determination from protein crosslink data \cite{ZhangJindal2024}. 

The ``vanilla'' EASAL algorithm first organizes an energy basin as a union of constant potential energy regions, and then treats each region, a distance-constrained configuration space, in geometric terms, as a \textit{branched covering space of a convex base space} represented in \emph{Cayley coordinates} (formally defined later) \cite{Ozkan2018ACMTOMS,Prabhu2020JCIM}. 
 Informally, Cayley parametrization 
 maps a topologically complex configurational region of effectively lower intrinsic dimension living in high dimensional ambient space  into a convex, and hence topologically simple, convex, Cayley base space \cite{Ozkan2018ACMTOMS,Prabhu2020JCIM} whose ambient dimension is the same as the lower intrinsic dimension.   Thus any type of  gradient descent and retraction maps are avoided. This allows all EASAL-based methods to bypass a major source of ill-conditioning and linearization errors, as well as inefficiencies in prevailing methods for enforcing constraints, including prevailing methods  that employ other types of internal coordinates that are not supported by the substantial  
Cayley parametrization theory. This theory, developed in \cite{SitharamGao2010,SitharamWilloughby2015,SitharamWang2014Beast,sitharam2018handbook}, helps address two major shortcomings shared by prevailing sampling methods including those that typically map between two coordinate systems (one ``internal'' and the other standard Cartesian). 

 While leveraging both above features of ``vanilla'' EASAL, the significant additional contribution of the new UC methodology, described in Section \ref{sec:algorithm_overview}, is sampling using a judicious combination of Cayley and Cartesian coordinates. This methodology is partially inspired by a slicing algorithm for 3D printing very large objects filled with mapped (curved) microstructures \cite{YOUNGQUIST2021103102}. See Figures \ref{fig:composite}, \ref{fig:cartesiancayley} and \ref{fig:curved_map}. As a by-product of the UC methodology, we develop a space-efficient traversal method for grid hypercubes of the ambient high energy configuration space that intersect the effectively lower dimensional, low, constant energy region of configuration space. The method is of independent interest and takes sublinear space on average in the number of grid hypercubes within a configurational region, both in a formal sense, and empirically. 
 
 All variants of the new  UC  methodology avoid matrix derivative (Jacobian and Hessian computations, subject to ill-conditioning and numerical errors) customarily used for conversion between internal (in this case Cayley) and Cartesian degrees of freedom. This is achieved by a   combination of three strategies that entirely avoid ill-conditioning errors and mitigate linearization errors: (i) back-and-forth between the low-computational-cost Cayley  parametrization and its inverse; (ii) utilizing an efficient geometric data-structure for keeping track of already sampled Cartesian grid hypercubes; and (iii)  flexible  heuristics - permitting tradeoffs between accuracy and efficiency - grounded in hypercube geometry that reduce the distortion caused by Cayley mapping of Cartesian hypercubes. This combination of Cartesian and Cayley coordinates yields a time-and-space-efficient sampling minimizing repetitions and discards. We also provide an open-source software implementation of several algorithmic variants of UC, built on top of EASAL 

The emphasis of this article is not extensive benchmark comparisons with multiple test cases and prevailing methods, nor to demonstrate large-scale, parallelized implementation. Rather, the emphasis here is on describing proof-of-concept results demonstrating the unique features of algorithmic variants of UC, which are based on rigorous geometric underpinnings, providing both proofs of efficiency and accuracy. However, we do demonstrate comparisons of the performance of the new methodology's algorithmic variants with the prevailing Monte Carlo based method, and, for completeness, also with ``vanilla'' EASAL \cite{OzkanBiCoB2011, Ozkan2018ACMTOMS, Prabhu2020JCIM} which was already shown to have certain specific performance advantages in comparison to prevailing Monte Carlo based methods in \cite{Ozkan2021JCTC, Zhang2022, ZhangJindal2024}. The new comparisons are performed via several computational experiments and measurements that highlight the new UC variants' accuracy of volume computation, sampling efficiency, coverage, and tradeoff between accuracy and computational resources. In-house computational experiments are performed on HiPerGator, University of Florida's supercomputing system. The best performing UC variant, called EASAL-hybrid-UC and described in detail in Section \ref{sec:step3} and \ref{sec:step3appendix}, promises to become both an effective stand-alone tool and as a complement to other widely used methodologies for computing configurational entropy. 

\subsection{Organization}
The rest of Section 1 summarizes related work  and prevailing methods. Section \ref{sec:methods} describes the new UC methodology, including the formal problem description, informal background on key techniques adopted from the ``vanilla'' EASAL methodology, and the details of the UC methodology to deterministically compute high-dimensional configurational (discrete) volumes for computing configurational entropy, as well as formally proven performance guarantees. Much of the rigorous mathematical treatment related to Section \ref{sec:methods} is relegated to the Appendix. Section 3 describes results of computational experiments including setup, key measurements, and various types of comparisons between new UC variants and Monte Carlo and, for completeness, also to the ``vanilla'' EASAL method. Section 4 discusses the UC variants' strengths and weaknesses and future work.

\subsection{Related Work}
\label{sec:related}
Even for simple systems of soft-matter assembly driven by relatively short-ranged pair-potentials \cite{Baker1999, WalesBogdan2006, Rybkin2013, BussiBranduardi2015, Li2023}, free energy computation using classical statistical mechanics methods continues to be researched, including for hard disks \cite{Ericok2022}, hard spheres \cite{Ericok2021}, and dumbbells (which resemble diatomic molecules) \cite{Hutter2022}. Such methods are also applied to larger systems as \cite{Chan2021} tabulating configurational entropy of more than 100 thousand small molecules with corrections from experimental data. 

For more complicated scenarios, such as biological macromolecules (proteins, for example), free energy computation remains an active research topic for more than 40 years \cite{Karplus1981, Go1983, Levy1984, Karplus1987}. Multiple methods have been attempted for different assembly-related processes involving proteins, including folding \cite{Hao1994, Fogolari2015}, recognition \cite{Frederick2007}, docking \cite{Ruvinsky2007, Lill2011} of small molecules, protein-protein \cite{Chang2007, Sun2017, Qiu2018} and protein-ligand \cite{Gao2010, Harpole2011, Duan2016, Verteramo2019} binding, etc. 

The underlying challenge that has occupied computational chemists for decades is to go beyond simplified scoring functions \cite{Pearlman2001, Liu2017} in the computation of configurational entropy \cite{ZhouGilson2009}, which requires accurate calculations of the relative volume of basins and regions of configurational space, using only mainstream computational molecular science methods, such as Monte Carlo algorithm \cite{Hao1994, Cheluvaraja2004} and Molecular Dynamics \cite{Case2005}. In aforementioned scenarios, uniformly sampling and computing (relative) discrete volume measures of lower energy and effectively lower dimensional and topologically complex spaces of higher energy, higher dimensional ambient spaces are crucial tasks. The ambient dimension is $6(k-1)$ when assembling $k$ rigid molecular components. Due to the topological complexity and dimensionality, prevailing stochastic methods are not generally ergodic, requiring multiple trajectories starting from multiple initial configurations.

The sources of topological complexity are two-fold: first, arising from both the structure of an energy basin with barriers, narrow channels etc., i.e. a network of nonlinear configurational regions at different energy levels and effectively different intrinsic dimensions (especially arising from short-ranged pair-potentials); and second, the topological complexity of each contiguous constant-energy region within the basin. All prevailing methods are afflicted by these twin ``curses of dimensionality and topological complexity''.

Other than EASAL \cite{Prabhu2020JCIM}, only a few methods (such as \cite{Holmes-Cerfon2013}) discern the difference between the above mentioned two types of topological complexity.
The first type of topological complexity necessitates extraction of a comprehensive \emph{roadmap} of a potential energy basin prior to intensive sampling, that is, the neighborhood and boundary relationships between constant energy regions within the basin at different energy levels, especially when they feed into multiple basins. This is best represented as a directed acyclic graph, with nodes representing the regions and directed edges representing neighborhood or boundary relationships with direction pointing from higher to lower energy levels. {\sl A more detailed definition of roadmap is provided in Section \ref{sec:background}, with full mathematical formality in \ref{sec:preliminariesappendix} and \ref{sec:previousappendix}}.

A few recent heuristic methods infer a roadmap \cite{Gfeller2007, Varadhan2006, Lai2009, PradaGracia2009, Yao2009}, starting from Monte Carlo and Molecular Dynamics trajectories, but only as a by-product of the full-blown sampling process. 
In the robotics motion planning literature, which deals with a version of an assembly landscape, there are exponential time algorithms to compute a roadmap (a version of atlas) and paths in general semi-algebraic sets, \cite{Canny1987, Canny1993, Basu2000} with probabilistic versions to improve efficiency \cite{kavraki1996, kavraki1998}. A geometric rigidity approach was primarily used to characterize the graph of contacts of arbitrarily large jammed sphere configurations in a bounded region. \cite{kahle2012, donev2004}. 
For the Cartesian configurational regions of nonintersecting spheres, the works \cite{baryshnikov2014, kahle2011} characterize the complete homology, viable only for relatively small point sets or spheres, while more empirical computational approaches for larger sets \cite{carlsson2012, bubenik2010}come without formal algorithmic guarantees. 
While the work \cite{Jaillet2017} also uses the word ``atlas'' (as we do in our EASAL methodology, described in Section \ref{sec:previous}), and refers to a configurational region stitched together from custom-sampled subregions called ``charts'', their subregions are not constant potential energy regions, and their organization into a roadmap are entirely unrelated to the EASAL methodology. They do not focus on assembly systems, but flexible systems such as cyclo-octane, where loop-closure type constraints predominate. The paper \cite{Holmes-Cerfon2013} uses a partition into constant potential energy regions that they call “modes,” and formally showed that their (and our methodology’s) geometrization of short-ranged pair-potential systems is physically realistic. However, they resort to prevailing methods such as Monte Carlo to compute configurational entropy or volumes of higher dimensional regions. 

In summary, most prevailing methods do not extract a roadmap prior to intensive sampling for configurational entropy or free energy computation. For example,  prevailing methods are stochastic and rely on random-walks in the high dimensional ambient space, and are not in general ergodic, i.e. cannot guarantee coverage of regions within a basin separated by barriers or narrow lower energy channels of effectively lower dimension, which arise in the presence of short-ranged pair potentials which are essentially distance constraints. Hence they require unpredictably long trajectories starting from many different initial configurations or microstates for locating such nearly separated regions. Methods for handling broken ergodicity include basin-hopping \cite{wales2013}, the use of parallelism \cite{earl2005}, or some combination of the two \cite{griffiths2019, wales2018}. All of these methods are extremely resource intensive and rely on inefficiently heavy sampling, which makes accurate computation of configurational entropy and free energy inefficient and impractical even with today's top-of-the-notch supercomputers.

While typical Molecular Dynamics based methods also do not extract a roadmap prior to sampling, they recognize the second type of topological complexity, namely of constrained constant potential energy regions of lower energy and lower intrinsic dimension. They attempt to work within the lower energy level or lower dimension using two techniques: 
\begin{itemize}
    \item by converting between two coordinate systems, one of which being the ambient Cartesian and the other being some form of internal coordinates \cite{Schwieters2001}, and
    \item relying on gradient descent and retraction maps in high ambient dimension to enforce these constraints and any additional restraints. 
\end{itemize}
For conversion between two coordinate systems, most of these methods use matrix derivatives - Jacobian and Hessian - of the mapping function to calculate steps to traverse the feasible region of configurational space and calculate volume \cite{Tribello2019}. Various ways of choosing internal coordinates are attempted, including natural internal coordinate \cite{Pulay1979, Fogarasi1992}, redundant internal coordinate \cite{Pulay1992, Peng1996}, delocalized internal coordinate \cite{Baker1996, Baker1999} etc. All these methods fall short for requiring computationally expensive pseudo inverses, such as the Moore-Penrose, which leads towards both high time cost and potential errors in calculation. 

More recently, the paper \cite{Rybkin2013} reported using higher order derivatives and their Taylor expansion to calculate steps for the sampling procedure and performed more efficient and reliable than traditional methods, and \cite{Oenen2024} focused on a coordinate system specified towards molecular vibrations. Software level optimization on the transformation between Cartesian and internal coordinates is also performed \cite{Wang2016, Bayati2020}, including involving machine learning-based predictions \cite{Li2023}. However, problems that arise from the nature of these methods still remain. Namely, linearization in numerical calculation creates error, and it will in turn harm the quality of the traversal of configurational space; also, ill-conditioning of the mapping function for some region of the configurational space will also greatly hinder the viability of using such methods for volume calculation, and entropy as well \cite{Demmel1990}. 
 
For staying within a constrained, lower energy or effectively lower dimensional, nearly constant potential energy region within a higher ambient dimension confounds most prevailing methods, forcing them to rely on local linearization and energy gradient descent (to enforce the constraints) leading to many discarded samples. For example the paper \cite{Holmes-Cerfon2013} needs to solve a nonlinear equation iteratively when tracing a 1 or 2-dimensional constrained manifold by moving a Cartesian configuration along its tangential direction and projecting it back to the manifold. Stratified sampling \cite{Dinner2017} could be considered a general way to address this problem by using stratified probability distributions whose support is restricted to a constant potential energy region. However, membership of a configuration in a region is characterized by the tangent space (linearization using Eigenvectors) at that particular configuration. Hence, the configuration needs to be sampled first, which does not address the original difficulty of staying primarily within the bounds of the desired region, that is, computing the bounds of the support of the probability distribution. In general, while the above-mentioned methods may address the accuracy issue in sampling low constant energy, effectively low dimensional regions, they do so at the expense of many discarded samples and efficiency.


\section{Methods}
\label{sec:methods}
In this section we describe our main contribution, the  UC methodology,  significantly extending and leveraging the EASAL methodology \cite{SitharamGao2010, Ozkan2018ACMTOMS, Prabhu2020JCIM, Wu2020PLoS} and open source software implementation \cite{Ozkan2018ACMTOMS, Prabhu2020JCIM, Wu2020PLoS}. 

\subsection{Background}
\label{sec:background}

For a precise description of the problem solved by the UC methodology, we start with a brief mathematical description of the overall assembly setup using distance constraints in Section \ref{sec:preliminaries}. Then, in Section \ref{sec:previous}, we briefly cover the underlying ideas and techniques from the basic EASAL methodology, followed, in Section \ref{sec:algorithm_overview} by two key problems that we tackle here, and corresponding challenges. Sections \ref{sec:step3},\ref{sec:step2} then informally describe the ingredients of the new UC methodology and its algorithmic variants' approach to solve these two problems. Much of the formal, mathematical treatment is relegated to \ref{sec:step3appendix} and \ref{sec:step2appendix}.

\subsubsection{Preliminaries: Problem Setup}
\label{sec:preliminaries}
We start by giving a brief formalization of assembly using distance constraints.
An \textit{assembly constraint system} $C$ is specified by:
\begin{itemize}
    \item A finite set $S = \{S_1, S_2, ..., S_k\}$ of $k$ finite point sets in 3 dimensional Euclidean space.
    \item A set of distance (interval) constraints between points in different $S_i$; these represent the pair potentials, see below.
\end{itemize}
The variables are the \textit{Euclidean/Cartesian} isometries $T_{S_i}$ for each $S_i \in S$, given by 6 scalars specifying $S_i$'s orientation (translation and rotation) relative to some fixed $O \in S$, where $T_O$ is assumed to be the identity.

Constraints of the system are derived from a discretized version of so-called Lennard-Jones potential constraint between pairs of atoms in assembling rigid molecules \cite{SEO2018}. Such potentials encode a variety of types of weak interactions, including Van der Waals, hydrogen bonds, as well as electrostatic, hydrophobic, hydrophilic, and quantum-level interactions. For our intended application, each of the constraints belongs to one of the following categories:
\begin{itemize}
\item $C_1$: For every $A \in S, B \in S, A \neq B,$ and for every $ (a, b), a \in A, b \in B$, $\lVert T_A(a)-T_B(b) \rVert \ge l(a,b)$. That is, every pair of points belonging to different point sets has a lower bound on the distance between them. Intuitively, these constraints represent a huge energy barrier and/or prevent collision between solid particles (atoms, residues, etc). 
\item $C_2$: There is at least one  $(a,b)$ where $a \in A, b \in B, (A, B)\in S$, $\lVert T_{A}(a)-T_{B}(b) \rVert \le h(a,b)$. In other words, some pairs are additionally constrained by distance upper bounds, and combined with $C_1$, by distance intervals. Constraints in this group strongly discourage the constrained point pair from being arbitrarily far away from each other where their interaction would cease.
\end{itemize}
While the formalization readily adapts to further refinement of constraints of the type $C_2$, such refinements only reduce the core challenges of configurational entropy or configurational region volume computation as we discuss below. 

The \textit{configuration space} 
is the set of configurations 
such that all constraints in $C_1$ and $C_2$ are satisfied. Our goal in this article is to efficiently and accurately calculate the (relative) volume of relevant \textit{configurational regions} or subsets of the configuration space. Furthermore, although wider Lennard-Jones wells with larger $(h-l)$ or with a graded staircase structure of the well can be combined with the EASAL methodology as described in detail in \cite{ZhangJindal2024}, our focus in this article is specifically to deal with the core difficulties posed by the limiting case where the interval size $(h-l)$ tends to 0. 
In this case, the configurational region satisfying one constraint of type $C_2$ is a nonlinear space of one lower energy level and dimension \cite{graver1993combinatorial, sitharam2018handbook} than the ambient space satisfying only $C_1$, whose energy level is proportional to the number of degrees of freedom, or dimension, which is $m=6(k-1)$.
We ensure that each constraint of type $C_2$ being satisfied in this limiting case is \emph{independent} (see Figure 7 in \cite{Prabhu2020JCIM}) and guaranteed to reduce the energy level and dimension of the configuration by one using generic combinatorial rigidity principles  \cite{graver1993combinatorial, sitharam2018handbook}, so we deal with a space of lower dimension than the ambient. A collection of ``small'' interval constraints generically yield a configuration space that is still of ambient dimension, but is a ``thin'' sheet (whose thickness tends to 0 in the limiting case) that is effectively lower dimensional. See Figure \ref{fig:cartesiancayley}.

Solving such a system requires traversal of the configuration space, which is commonly done by traversing the ambient space while trying to stay within the lower energy, effectively lower dimensional, topologically complex region, i.e., which has disconnected components, holes, handles, narrow channels etc. These difficulties are compounded several fold in the limiting case where $(h-l)$ is close to 0, which motivates our treatment here. See Figure \ref{fig:cartesiancayley} and similar EASAL screenshots in \cite{Prabhu2020JCIM}. As mentioned in Section \ref{sec:related}, gradient descent, which involves alternating linear tangential steps and corrections back to the configuration space, is commonly used in prevailing methods including molecular dynamics and Monte Carlo based methods. This pervades many common tasks such as finding optimal or extremal configurations, finding paths, path lengths, region volumes (configurational entropy), path probabilities for transition networks, sampling configurations or paths etc. Such traversal is impractical when the number of points in the point sets and number of distance interval constraints in $C_2$ is large. We simultaneously face the combined curses of ambient dimensionality and topological complexity of the constrained region, calling for a novel approach for mitigating such issues.
\begin{figure}
\centering
    \includegraphics[width=.9\textwidth]{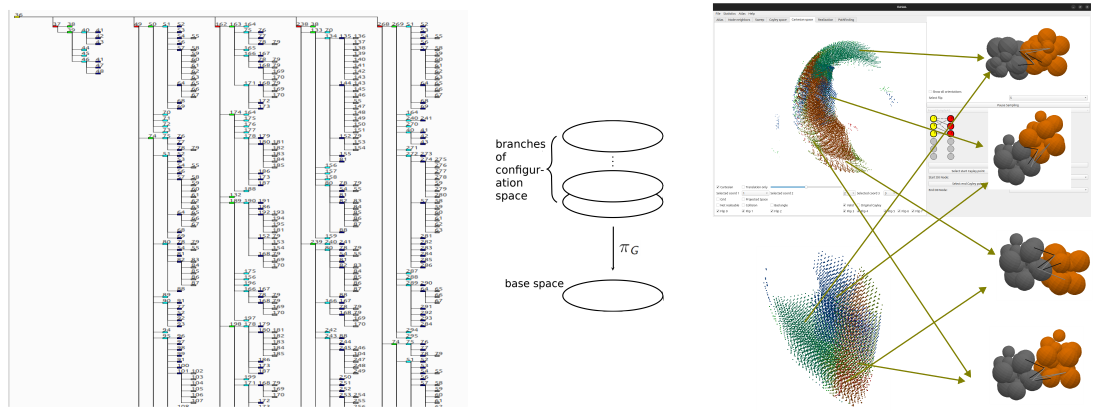}
    \caption{Vanilla EASAL methodology}
    \scriptsize
     See Section \ref{sec:previous}.
    From left to right: 
    \textbf{1st:} EASAL screenshot of part of the roadmap DAG  of the configuration space of  assembly system shown at far right. DAG is shown as a tree for convenience, with each node representing a contiguous constant potential energy region or active constraint region (ACR); with arrows pointing from higher to lower energy or effective dimension
    \textbf{2nd:}  Schematic  of a covering map, namely the Cayley parametrization showing branches (called branches or \emph{flips}) of the covering space and corresponding base space.  The branches or flips represent chirality classes.
    \textbf{3rd:} Screenshot of EASAL showing one ACR uniformly sampled in Cayley (below) and therefore highly non-uniformly sampled in Cartesian (above): an intrinsically 3-dimensional covering (Cartesian) space representing the ACR, i.e. satisfying 3 constraints of type $C_2$ and hence of energy level 3 steps below its  6 dimensional ambient space, which satisfies only constraints of type $C_1$;  and its simple base (Cayley configuration) space whose 3 intrinsic and ambient dimensions are the 3 Cayley coordinates;  colors represent different branches/flips
    \textbf{4th:}  EASAL screenshot of selected configurations in different  branches/flips of  the indicated ACR.
    \label{fig:composite}
\end{figure}
\subsubsection{EASAL Background: Roadmapping Basin Regions, Cayley parametrization}
\label{sec:previous}
The papers \cite{Prabhu2020JCIM,Ozkan2018ACMTOMS,OzkanBiCoB2011} solved the problem of \emph{deterministically} traversing or sampling -- while staying within -- a Cartesian configuration space constrained by $C$, as described above, using strategies including:
\begin{enumerate}
    \item Generating an accurate atlas of configuration space with minimal sampling, including a roadmap of stratified neighborhood relationships between constant energy regions, each of which satisfies a different subset of $C_2$, called \textit{active constraint regions} (ACRs) and defined below. In other words, roadmap generation (configuration space topology extraction) is decoupled from sampling (for each ACR). In particular, the methodology uses a  energy level or effective dimensional stratification in the atlas using combinatorial rigidity principles \cite{graver1993combinatorial, sitharam2018handbook}: it starts from ambient high energy or high-dimensional interior (overall parent region) and recursively locates boundaries (child regions) of one less effective dimension (one lower energy level), which then become parent regions for their boundary child regions etc., a geometrically efficient task since the energy level or effective dimension decreases only by 1 each time. The atlas directly provides a decomposition of energy basins into constant energy regions stratified by energy level as described in detail below.
    \item Each ACR is sampled efficiently by avoiding any form of gradient descent or retraction maps to enforce constraints, which minimizes repeating or rejected samples. This is done by Cayley parametrization \cite{SitharamGao2010, SitharamWang2014Beast,Wang2014Caymos,SitharamWang2011CayleyI,SitharamWang2011CayleyII, SitharamWilloughby2015}, a distance-based internal coordinate representation of configurations within ACRs, discussed in detail below, which is particularly suited to ACRs which are entirely defined using distance based constraints such as pair potentials.
    Cayley parametrization essentially ignores the ambient dimension of ACRs for the aforementioned limiting case of $(h-l)$ being close to 0, where an ACR of lower energy level than the ambient is a topologically complex manifold (technically a semi-algebraic set) of effectively lower intrinsic dimension in higher-dimensional ambient space. EASAL ``flattens'' the lower, constant energy region into its own ambient space of dimension equal to its effective intrinsic dimension, as detailed below, making the sampling computationally practical and overcoming the curse of dimensionality. On the contrary, many conventional sampling methods, including Monte Carlo, sample such a region directly in high dimensional ambient space by stepping in tangent space, and using gradient descent, projections and other retraction maps to bring the samples back to the ACR, whereby they ignore or at least do not leverage the lower energy ACR's effectively lower intrinsic dimension.
    \item A significant simplification of traversal for sampling, and further minimization of repeated and rejected samples is achieved by convexifying ACRs \cite{SitharamGao2010}, again using Cayley parametrization, as detailed below and in \ref{sec:previousappendix}.
\end{enumerate}

We give a formal description of each of the above points in the following paragraphs and in the Appendix, then provide a brief outline of using EASAL to calculate volume of configuration space.

\medskip
\noindent\textbf{Roadmap and Basins}
The overall problem of satisfying constraints in the above type of assembly constraint system $C$ can be divided into subproblems with systems associated with $C$ as follows. Notice that the existential quantifier in $C_2$ can be replaced by a disjunctive union over all simultaneously satisfied subsets $Q \subseteq C_2$. Consider a variant of the system consisting of all collision-avoidance constraints in $C_1$, but replacing $C_2$ by the conjunction of distance upper bound constraints in a subset $Q$, i.e., the pairs in $Q$ represent those pairs that are in a Lennard-Jones well. Define the \textit{Active Constraint Region(ACR)} as the set of all configurations satisfying all constraints in $C_1$ and $Q$, then the original configuration space \cite{OzkanBiCoB2011,Ozkan2018ACMTOMS, Prabhu2020JCIM} is partitioned into a disjoint union of ACRs. The set of ACRs is stratified by the size of $Q$ and organized as a partial order (in the form of a \emph{directed acyclic graph (DAG)}) which we call the (topological) \textit{roadmap}. Each node of the roadmap DAG represents an ACR where all configurations have the same energy
 If we further assume that all the Lennard-Jones potential wells are of equal (without loss of generality, unit) depth, the level of stratification of an ACR, corresponding to a set of constrained pairs $Q$, is the relative energy of its configurations, whose proxy is its effective dimension, since each constraint in $Q$ reduces the energy level and effective dimension by 1 when close to the limiting case. 
 Directed edges between ACRs (or parent-child relationships), reflect the superset-subset relationship of the ACRs (which is also an interior-boundary relationship for the limiting case of extremely narrow Lennard-Jones wells), or alternatively the subset-superset relationship of the corresponding sets of constrained atom pairs $Q$. Additionally, since each constrained pair in $Q$ corresponds to one atom pair being in a discretized Lennard-Jones potential well, each ACR in the roadmap also represents a contiguous configurational region with constant energy level.

Both the DAG and stratification structures of the roadmap are used for exploration by the algorithm outlined below. This decouples the exploration of the roadmap from sampling process of each node in the roadmap.

\begin{figure}
\centering
    \centering
    \includegraphics[width=\textwidth]{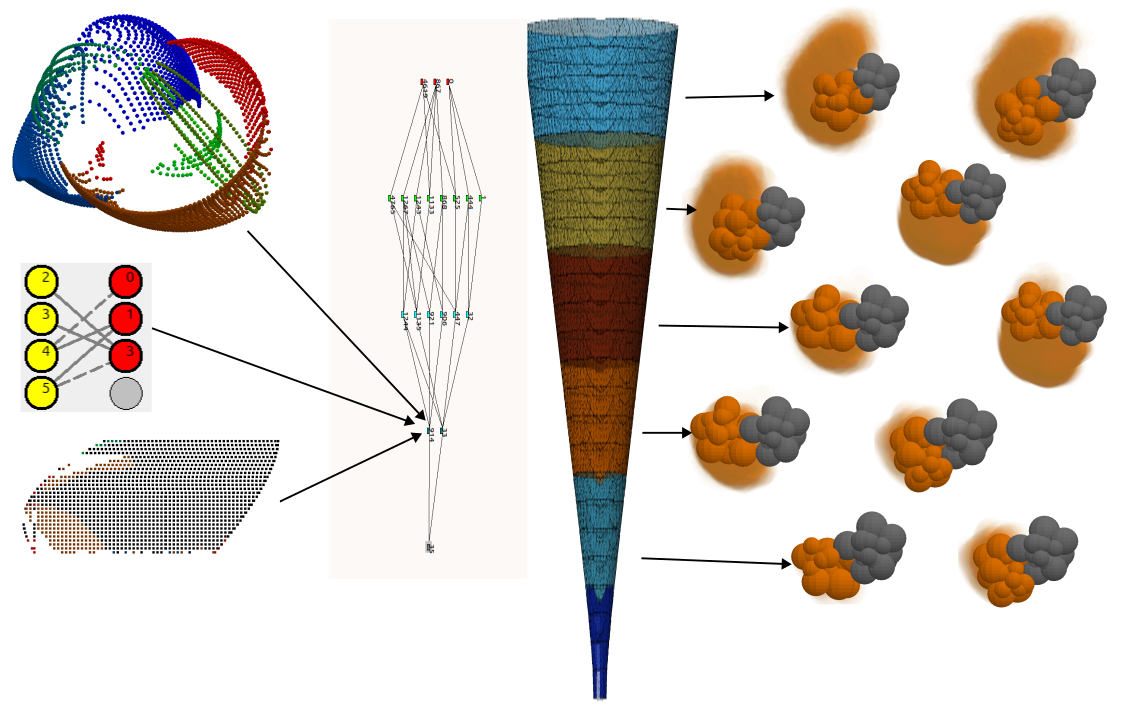}
\caption{Views/Components of Basin Structure, and Experimental Assembly system}
\scriptsize See Sections \ref{sec:background}, \ref{sec:previous}, \ref{sec:preliminaries}, experimental setup in Section \ref{sec:setup}, and \ref{sec:guide}. Illustration of different views and components of an energy basin. 
\textbf{Left:} EASAL screenshot of ACR that is non-uniformly sampled as Cartesian branched covering space (top), with branches or flips in different colors, and uniformly sampled in Cayley base space (bottom) (vanilla EASAL methodology); ACR is 4 energy levels below the  ambient (not shown) and is of effective dimension 2, indicated in the roadmap to its right and represented by the ACG showing 4 constrained pairs (bold edges) in the Lennard-Jones wells (left middle), and 2 Cayley parameters (dashed non-edges); \textbf{Mid-left:} EASAL screenshot of the basin DAG subset of the entire roadmap DAG; \textbf{Mid-right:} Basin schematic illustrating energy levels and their expected configurational volumes;
\textbf{Right:} At each energy level of the basin, EASAL screenshots of configurational sweeps of selected branches/flips of selected ACRs at the indicated energy level, for the 2 point sets in the experimental assembly constraint system $C$: $S_1$ (gray) and $S_2$ (orange). 
\label{fig:basin}
\end{figure}

A \emph{basin} can then be defined as a collection of ACRs, with a bottom energy level ACR having a maximal set of \emph{independent} satisfied constraints $Q$ and its upward closed set (i.e. all ACRs corresponding to sets $Q' \subseteq Q$ of constrained pairs). There is a direct, barrier-less, energy lowering path from any ACR in the basin to the bottom ACR. Note that different basins are not necessarily disjoint sets and can share ACRs. Different views of an energy basin are shown in Figure \ref{fig:basin}.

\medskip\noindent\textbf {Cayley Parametrization}
  Each ACR,  defined by a given set of constraint $Q$,   is of lower (constant) energy level or lower effective dimension than ambient $m=6(k-1)$ in the limiting case. Its  topologically complexity (see Figure \ref{fig:composite}) arises from the combination of $C_1$ and $Q$. Cayley parametrization \cite{SitharamGao2010, SitharamWilloughby2015,Ozkan2018ACMTOMS, Prabhu2020JCIM, Wu2020PLoS} is broadly applicable \cite{SitharamWang2014Beast, Wang2015, Sadjadi2021}, and maps the ACR into a convex, and hence topologically simple base space of the same dimension, consisting of Cayley configurations. In the limiting case where the region is ``thin'', Cayley parametrization further ``flattens'' it into an effectively $(m-\lvert Q \rvert)$ dimensional region residing in ambient space of the same dimensionality. 

Cayley parametrization is further motivated by the following desirable properties. The combination of these properties ensures high sampling efficiency and accuracy, avoiding gradient-descent or retraction maps to enforce constraints, avoiding repeated sample configurations and minimizing discarded sample configurations.
\begin{itemize}
    \item The Cayley parametrization is a \emph{covering} map from Cartesian ACR, called a \emph{branched covering space} onto the convex \emph{base} space which is computationally easy (constant time), essentially just measuring distances within a Cartesian configuration $T$ which have not been explicitly constrained.
    \item The boundaries of the convex Cayley base space are easy to compute, and depend entirely on the distance interval constraints in $Q$. Additional collision avoidance constraints in $C_1$ carve out typically a small number of convex Cayley parametrizable regions from this convex Cayley base space.
    \item The inverse (from Cayley to Cartesian) maps each Cayley configuration in the base space to generically finitely many Cartesian configurations  each uniquely identifiable by chirality, and easily computable (constant time). The collection of pre-image Cartesian configurations, a branch of the branched covering space containing configurations   of the same chirality is called a \emph{flip}.
\end{itemize}
Using these properties, one can efficiently traverse the lower, constant energy ACR, an effectively lower dimensional and topologically complex  Cartesian ACR, living in high ambient dimension as follows. Traversing the effectively lower dimensional convex base space of the ACR is efficient by (1-3) above; computing the pre-image configurations of the covering map is efficient by (3) above. This yields a traversal that does not leave the branched covering space, namely the ACR itself.

We now formally define Cayley parametrization \cite{SitharamGao2010, SitharamWilloughby2015, OzkanBiCoB2011, Ozkan2018ACMTOMS, Prabhu2020JCIM} in the context used for this article. 

Each ACR has an underlying \emph{active constraint graph (ACG)} $G_Q= (V_Q, E \cup E_Q)$, whose vertex set $V_Q$ represents the set of the points involved in the constraints in $Q$. There are two sets of edges:
\begin{itemize}
    \item $E$: all $(a, b)$ where $a$ and $b$ belong to the same point set in $S$.
    \item $E_Q$: pairs $(a, b)\in Q$.
\end{itemize}

Define a \textit{nonedge} of a graph as a vertex pair not in its edge set. One way to represent or parametrize a Cartesian configuration $T$ is using the tuple of distances/lengths attained in the configuration by a subset $F$ of nonedges of its ACG $G_Q$. This tuple is the corresponding Cayley configuration, with each Cayley coordinate or parameter (value) of the tuple being a nonedge length (value). Formally, this defines an easily computable Cayley parametrization map that maps the ACR into the Cayley base space. By choosing a minimal nonedge set $F$ generically of size $(m-\lvert Q \rvert)$ whose addition makes the ACG $G_Q$ rigid \cite{sitharam2018handbook}, we ensure that a Cayley configuration corresponds to at most finitely many Cartesian configurations. 
In \ref{sec:previousappendix}, we formally define \emph{nice} ACG's $G_Q$ (as characterized in the papers\cite{SitharamGao2010,Prabhu2020JCIM}) that ensure that this Cayley parametrization map additionally satisfies the above mentioned properties of convex base (image) space and easily computable inverse. We note that in the case of $k=2$ assembling components (point sets in the formal problem description above), almost all (over 85\%) of the ACGs are nice.

\subsection{Basin Configurational Volume: Formal Computational Problems Addressed by New Methodology }
\label{sec:problem_description}
 For free energy computation,  a good approximation of our scenario is that each satisfied constraint in $C_2$ reduces energy by the same amount, then the appropriate configurational entropy measure for a basin is the weighted volume, obtained by equally weighting volumes of the ACRs at each dimension or energy level within the basin. Specifically, given a Boltzmann factor $B$ associated with each energy level, the weighted volume of a basin is $V=\sum_{k=1}^{m-1}B^{(m-1-k)} V_k$, where $V_k$ is the summation of (unweighted) volumes of all $k$-dimensional ACRs in the basin, which corresponds to a constant energy level that is $(m-k)$ energy steps below the ambient.
To calculate energy basin volumes, it is necessary to accurately and efficiently calculate the Cartesian volume of each and every such constant energy ACR at every energy level in the basin. While EASAL's Cayley coordinate representation provides multiple benefits \cite{SitharamWang2011CayleyI,SitharamWang2011CayleyII,Wang2014Caymos,SitharamWang2014Beast,Ozkan2018ACMTOMS,Prabhu2020JCIM,Ozkan2021JCTC,Wu2020PLoS,ZhangJindal2024}, it is not clear how to use it to sample the configuration space \emph{uniformly} in the original \textit{Cartesian} coordinates of the ambient space or to compute volume measures \cite{Hikiri2016,Gyimesi2017ConfigurationalEntropyGaussianMixture,Fogolari2015}. To calculate volume defined based on a Cartesian coordinate grid, we need to solve the following problems:

\noindent
\textbf{Problem 1} is to compute the $\epsilon$-approximate \textit{volume} of a (feasible) configuration space in ambient $m=6(k-1)$-dimensional Cartesian coordinate space, defined as the relative proportion of $m$-dimensional hypercubes of side length $\epsilon$ that intersect the region (in a generic rectilinear grid subdivision of the ambient space). 

\noindent
\textbf{Problem 2} is to generate one point in the region per intersecting hypercube as in Problem 1, i.e. a uniform Cartesian sampling of configurations in the ACR. 

Both problems assume distance ACGs $G_Q$ with constraint set $Q$ of type $C_2$ in the nice class introduced in Section \ref{sec:previous} and defined formally in \ref{sec:previousappendix}.
Clearly Problem 1 reduces to Problem 2, but it is conceivable that Problem 1 could also be solved directly. As we explain the approaches to Problem 2 below, we note the reasons why, in current practice, Problem 1 is solved essentially by solving Problem 2.

\begin{figure}
\centering
\includegraphics[width=\textwidth]{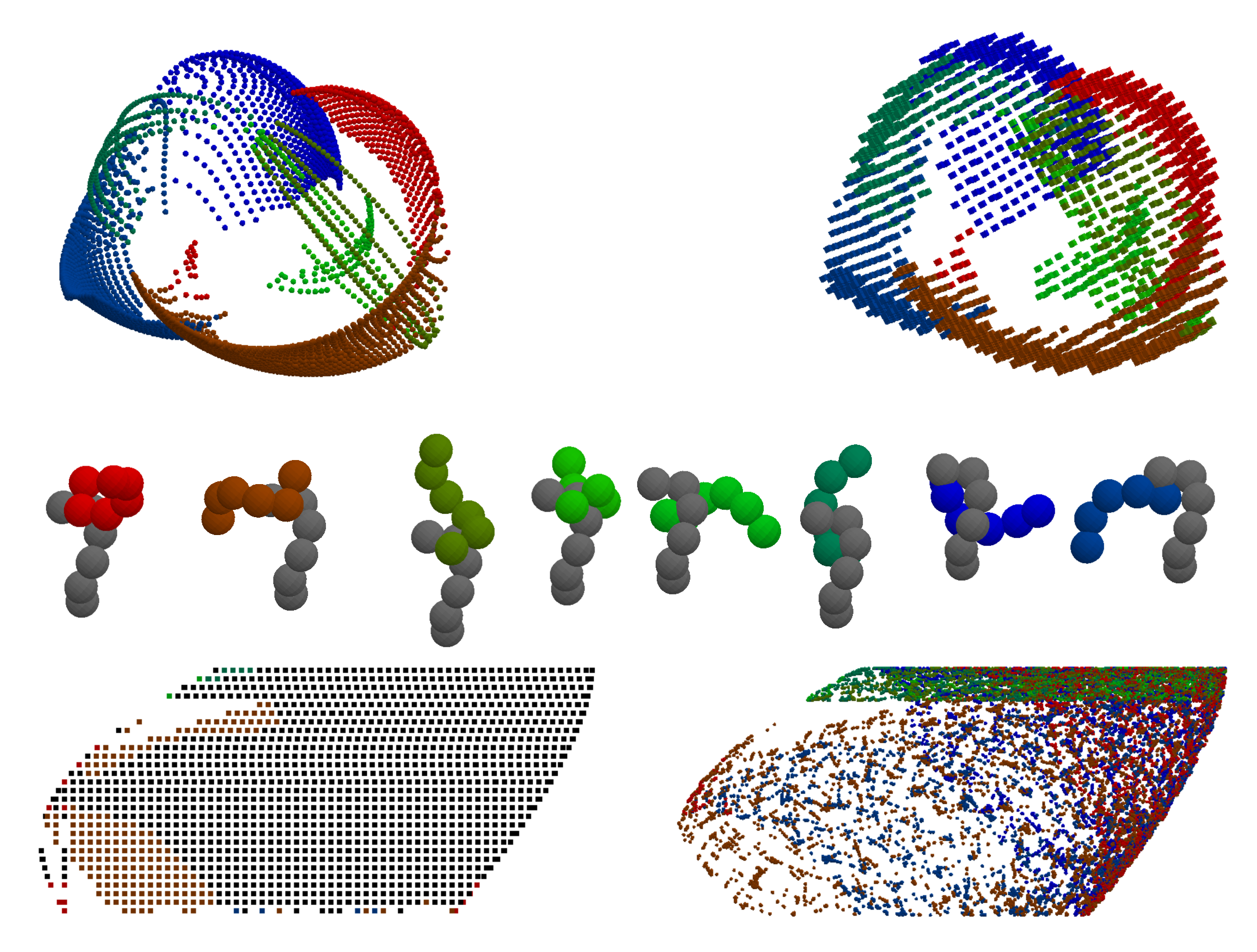}
\caption{(non)Uniformity in Cartesian and Cayley space}
\scriptsize See Sections \ref{sec:previous} and  \ref{sec:algorithm_overview}. \textbf{Top-left:} vanilla EASAL and UC screenshots of 2-dimensional (4 energy levels below the ambient) Cartesian configuration space (of a distance constraint system in dimension 3) in 6-dimensional ambient space of highest energy level, projected on 3 translational coordinates, non-uniformly sampled; colors represent different branches/flips of the configuration space viewed as a branched covering space. 
\textbf{Middle:} some feasible configurations, sampled by vanilla EASAL  and  UC, of point sets of toy assembly system (\emph{for illustration purposes only -   not  used for experiments}) satisfying all the constraints, one from each branch/flip or chirality class. 
\textbf{Bottom-left:} corresponding convex base space with uniform sampling in Cayley coordinates. Note that some Cayley configurations have multiple colors - each such configuration has finitely many pre-images each belonging to a different flip. 
\textbf{Top-right:} Uniform Cartesian sampling of a branched covering space (colors represent different flips or branches) of another 2-dimensional configuration space (4 energy levels below the 6-dimensional ambient space of highest energy level, projected on 3 translational coordinates). 
\textbf{Bottom-right:} Corresponding (non-uniform) sampled points in the convex base space in Cayley coordinates.
\label{fig:cartesiancayley}
\end{figure}

A direct way \textbf{(*)} to try to solve both problems is to use the known efficient method \cite{Ozkan2018ACMTOMS,Prabhu2020JCIM,Wang2014Caymos,SitharamWang2014Beast} for uniform sampling in Cayley coordinates of ACR's convex base space together with pre-image computations as described above. However, with no adjustments, this results in a highly nonuniform sampling of the ACR, i.e. an unsatisfactory solution to Problem 2, see Figure \ref{fig:cartesiancayley}.

In fact, this is a decades-old problem in computational chemistry, referred to as ``internal coordinate to Cartesian back transformation'', that continues to be actively studied \cite{Baker1999,Rybkin2013,Li2023}. To clarify, the ``internal coordinates'' used in computational chemistry, e.g. in molecular dynamics, are different from Cayley coordinates, and to the best of our knowledge lack the underlying theory and tools available for Cayley coordinates.

The straightforward workaround is to traverse the base space of the ACR in Cayley coordinates, but iteratively adjust the size and direction of each Cayley step by computing a pseudoinverse of the linearization (Jacobian) of the covering map and ensuring uniform Cartesian sampling of the pre-image branched covering space. This approach suffered from both inaccuracies due to linearization error as well as ill-conditioning problems, thus had overall unreliable performance on accuracy and efficiency \cite{Ozkan2014Jacobian}. A standard way to address these problems is to use the Hessian and higher derivatives of the covering map. However, such efforts are still underway\cite{Baker1999,Rybkin2013,Li2023} and the problem is by no means settled. 

It should be noted that the use of derivatives of the covering map theoretically provides another approach to Problem 1 directly without Problem 2 \cite{Dyer1991, applegate1991, kannan1997, LOVASZ2006, Ge2015AFA} by using the derivatives of the covering map to compute the volume of the configuration space and solve Problem 1. In any case, this too involves some form of randomized or deterministic adaptive sampling in Cayley coordinates. Furthermore, to our knowledge such an approach to Problem 1 that avoids Problem 2 does not exist in the literature. One reason could be that although the covering map is quite well behaved, this cannot be said about the pseudoinverses of its Jacobian or Hessian. Our contributions provide an optimal solution to Problem 2 and thereby a solution to Problem 1, see, e.g. Figure \ref{fig:composite} for an illustration of quality obtained.

\subsection{Details of the New Methodology}
\label{sec:algorithm_overview}
First, we reiterate that all EASAL-based methods use the same method for roadmapping basins. 
 Additionally, all EASAL-based methods  use Cayley parametrization to 
 map a topologically complex ACR  of effectively lower intrinsic dimension living in high dimensional ambient space  into a convex, and hence topologically simple, convex, Cayley base space whose ambient dimension is the same as the lower intrinsic dimension.   Thus any type of  gradient descent and retraction maps are avoided. This allows all EASAL-based methods to bypass a major source of ill-conditioning and linearization errors, as well as inefficiencies in prevailing methods for enforcing constraints, including prevailing methods  that employ other types of internal coordinates that are not supported by the substantial theory \cite{SitharamGao2010,SitharamWilloughby2015,SitharamWang2014Beast,sitharam2018handbook} that ensures the convex covering map property of Cayley coordinates.
 
The difference between the EASAL-based methods lies in whether they combine Cartesian with Cayley coordinates (``vanilla'' EASAL does not), and how the new, combination methods, collectively referred to as ``variants of \textit{Uniform Cartesian} (UC)'', perform intersections and search.

 Second, all variants of UC avoid matrix derivative (Jacobian and Hessian computations, subject to ill-conditioning and numerical errors) customarily used for conversion between internal (in this case Cayley) and Cartesian degrees of freedom. This is achieved by a judicious combination of three strategies that entirely avoid ill-conditioning errors and mitigate linearization errors: (i) back-and-forth between the low-computational-cost Cayley  parametrization and its inverse; (ii) utilizing an efficient geometric data-structure for keeping track of already sampled Cartesian grid hypercubes; and (iii) flexible  heuristics - permitting tradeoffs between accuracy and efficiency - grounded   in hypercube geometry that reduce the distortion caused by Cayley mapping of Cartesian hypercubes. This combination of Cartesian and Cayley coordinates yields a time-and-space-efficient sampling minimizing repetitions and discards.
 
Here we describe three new features of all variants of UC.

\begin{figure}[htbp]
    \centering
    \includegraphics[width=.7\linewidth]{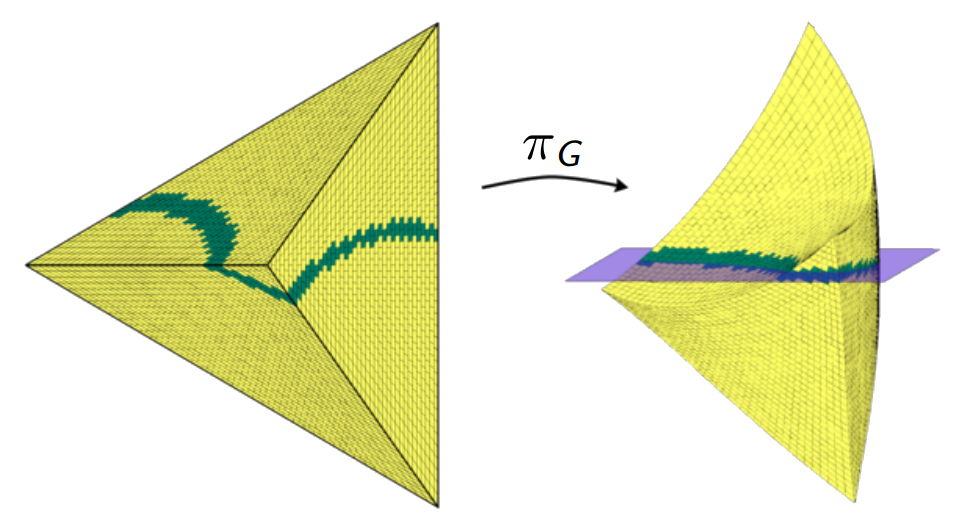}
  \caption{Illustration of intersection of ACR and grid hypercubes in Cartesian and Cayley. }
  \scriptsize Schematic, conceptual illustration of a 2-dimensional configuration space (green) in 3-dimensional ambient space (yellow), intersecting grid cubes, before and after Cartesian to Cayley covering map is applied (figure courtesy \cite{YOUNGQUIST2021103102}). Note that in Cayley, the intersection is with a linear subspace.
  Note that in this article, the ambient space is at least 6-dimensional when the number of point sets $\lvert S \rvert = k$ in the constraint system $C$ is exactly 2, as in the results Section \ref{sec:results}.
\label{fig:curved_map}
\end{figure}

\begin{enumerate}
    \item A principled, formally proven, solution to Problem 2 when the ACG $G_Q$ is in the aforementioned 
nice class in time linear in the output size, i.e. in the number of $\epsilon$-cubes that intersect the configuration space. A few natural variants of the intersection step illuminate the accuracy-efficiency tradeoffs. In addition to leveraging the known efficient method (*) for sampling the base space in Cayley coordinates \cite{Ozkan2018ACMTOMS,Prabhu2020JCIM}, the new methodology is partly inspired by a slicing algorithm for 3D printing very large objects filled with mapped (curved) microstructures \cite{YOUNGQUIST2021103102}. See Figures \ref{fig:composite}, \ref{fig:cartesiancayley}, and \ref{fig:curved_map}.

    \item A space-efficient grid traversal method (and natural variants) that on average takes sublinear space in the number of grid cubes visited. This indicates sublinear space complexity (in terms of output size) for  modified Problems 1 and 2 that require \textit{at least} (instead of exactly) one point per grid hypercube that intersects the ACR. 

    \item An opensource software implementation that is used to compare the performance of our method for Problem 1 using the obvious ``vanilla'' EASAL method (*) described above, i.e., Cayley sampling together with pre-image computations of the covering map, which was already shown to have significant advantages in efficiency and efficiency-accuracy tradeoffs, in comparison to prevailing Monte Carlo based methods in \cite{Ozkan2021JCTC}. The new UC implementation relies on efficient grid hypercube representations that could be of independent interest: they speed up the extraction of arbitrary dimensional facets and simplices and their intersection with the ACR.
\end{enumerate}

For completeness, we recall the \textbf{input} to the algorithm which includes the set $S$ of point sets, the constraints $C=(C_1, C_2)$ of Section \ref{sec:preliminaries}, ACR defined by the ACG $G_Q$, with the constraint set $Q$ of type $C_2$ in the nice class introduced in Section \ref{sec:previous} and formally defined in \ref{sec:previousappendix}, and the required accuracy $\epsilon$ for Problems 1 and 2.
 
For reasons of exposition and the current software implementation, we make several {\bf assumptions} that we justify as follows: we assume $\lvert S \rvert = 2$ and the ambient dimension $m = 6$. A roadmapping algorithm for $\lvert S \rvert > 2$ is given in \cite{Prabhu2020JCIM}, which can be used directly without altering the UC volume computation, but there are more ACGs that fall outside the ``nice'' class and have to be dealt with, e.g., in more traditional ways. We further assume the modified Problems 1 and 2 that require \textit{at least} (instead of exactly) one point per grid hypercube that intersects the ACR. From the solutions to these modified problems, a straightforward output data structure with a hash map yields solutions to the original problems.

All the algorithmic variants of UC have the 4 following steps.

\begin{enumerate}
\item Using the covering map (see Section \ref{sec:previous}, \ref{sec:previousappendix}, and \cite{SitharamGao2010}), \textbf{sample the base space} in Cayley coordinates, via the method in \cite{Ozkan2018ACMTOMS, Prabhu2020JCIM} that determines a Cayley step-size based on the volume discretization constant $\epsilon$ (see Problems 1 and 2 in Section \ref{sec:problem_description}) and finds boundaries and extremal configurations. Further compute the corresponding pre-image Cartesian configurations in the configuration space. 
\item For each flip, using the Cartesian $\epsilon$-grid hypercube containing configurations from the previous step as a starting point, \textbf{generate hypercubes on-demand, and traverse using a key frontier hypercube data structure, as described below}.
\item Use the covering map to generate the corresponding Cayley cuboid, then \textbf{calculate the intersection with $(m-\lvert Q \rvert)$-dimensional region satisfying all constraints in $Q$, as described below} but not necessarily the constraints in $C_1$; this generates partly feasible Cayley configurations $c$.
\item Compute the pre-image configurations, retain only if fully feasible, i.e. only if $C_1$ is satisfied, and \textbf{find and count the corresponding Cartesian grid cube} if the cube is in the flip. Note that due to linearization error, the original grid cube (created in step 2) may may differ from the corresponding grid cube. This solves the modified Problems 1 and 2. A straightforward output data structure stores the cubes and locates them with a hash map to avoid double counting. This solves Problems 1 and 2.
\end{enumerate}

We now describe Steps 2 and 3 in detail.

\subsection{Intersection Calculation, and Variants of UC in Step 3}
\label{sec:step3}

Step 3 in the overall algorithm is challenging due to the following reasons:
\begin{itemize}
    \item The map function is a non-linear (quadratic) map, thus the image for Cartesian cube in base space is a non-linear object, making direct intersection calculation unrealistic;
    \item The intersection yields a potentially disconnected $(m-\lvert Q \rvert)$-dimensional region without a tractable description.
\end{itemize}

To tackle both issues, we provide a series of workaround operations to calculate intersection:
\begin{enumerate}
    \item Instead of using a single, $m=6(k-1)$-dimensional cuboid intersection calculation, decompose each $\lvert m \rvert$-dimensional cube into a collection of $\lvert Q \rvert$-dimensional objects and map each component to Cayley space. 
    \item Linearize each decomposed component, which is also $\lvert Q \rvert$-dimensional.
    \item Calculate intersection between the mapped component and region satisfying $Q$ by solving a system of $\lvert Q \rvert$ linear equations for a convex combination.
\end{enumerate}
Taking advantage of co-dimension relationship between each component ($\lvert Q \rvert$-dimensional) and region satisfying $Q$ ($(m-\lvert Q \rvert)$-dimensional), their intersections are generically 0-dimensional, i.e. points. After these operations, each component generates at most one intersection point due to linearity of both entities. The set of intersection points generated by this process is then input into Step 4 of the overall algorithm.

\begin{figure}[htbp]
    \centering
    \includegraphics[width=.7\linewidth]{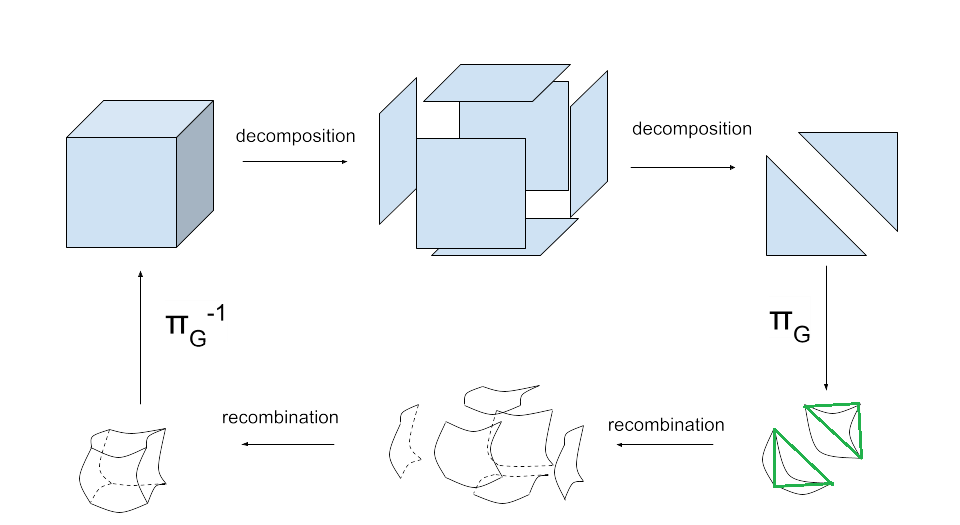}
    \caption{Illustration of hypercube decomposition}
    \scriptsize Schematic illustration of decomposition of a Cartesian hypercube into appropriate dimensional components, and linearized components of corresponding Cayley cuboid.
    \label{fig:decompose}
\end{figure}

\begin{figure}[htbp]
    \centering
    \includegraphics[width=.5\linewidth]{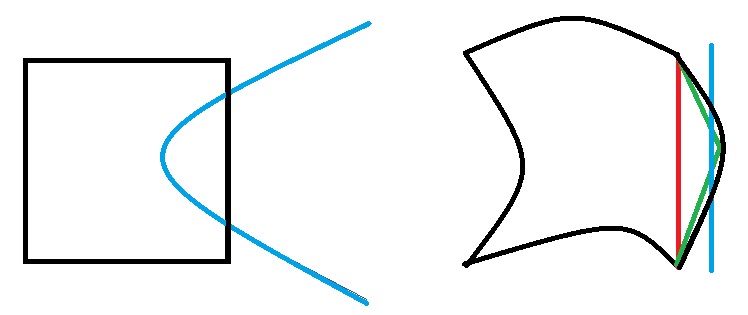}
    \caption{Simplicial decomposition error}
    \scriptsize Schematic illustration of error in EASAL-UC and how EASAL-face-UC fixes it. \textbf{Left:} hypercube(black) and ACR(blue) in Cartesian space, \textbf{Right:} hypercube(black) and ACR(blue)'s image in Cayley, EASAL-UC linearizes the component as red and EASAL-face-UC as green. Modified strategy helps finding intersections that are otherwise missed with regular linearization. 
    \label{fig:dogchewerror}
\end{figure}

 Algorithmic variants of UC provide several different ways of mapping, decomposing, and linearizing $p$ into the set of components for varying degrees of accuracy and efficiency in the intersection calculation. This includes a simplicial element-based decomposition called EASAL-UC (see Figure \ref{fig:decompose}), a modification on EASAL-UC (with notable improvement on accuracy, see Figure \ref{fig:dogchewerror}) called EASAL-face-UC, a hybrid version of the above 2 methods called EASAL-hybrid-UC, a fast yet coarse facet parallelepiped element based decomposition called EASAL-basis-UC, and a special case where we loosen $C_2$ to be distance interval constraints called EASAL-thick-UC. Among the methods, EASAL-hybrid-UC turns out to be the best performing method overall, while EASAL-thick-UC provides a straightforward solution to the more general case if $\lvert Q \rvert = 1$ is enforced. 
Please see \ref{sec:step3appendix} for further details.

\subsection{The Frontier Hypercube Graph Data Structure: Step 2}
\label{sec:step2}
To avoid repetition during decomposition and traversal of hypercubes, we designed a sophisticated data structure, called the \textit{Frontier Hypercube Graph}, to store visited hypercubes as well as their relationship with yet-to-visit hypercubes. 

As we decompose each hypercube into elements residing on the surfaces, found intersections can also be associated with those elements shared between the original hypercube and its neighbors. Therefore, during the traversal process, we can opt to only visit a neighbor if it shares an element with the current hypercube with a feasible intersection point found in it. This is accomplished by 2 hypercube containers, one for \textit{promising} hypercubes (those neighboring a visited hypercube, not yet visited, and share such feasible element), and the other for \textbf{not-yet-promising} (those neighboring a visited hypercube and do not share feasible element). The traversal is complete when the promising container is empty. \textit{The key property of this data structure is that a hypercube can be removed as soon as it is processed without compromising the traversal}. By using this approach the space complexity of UC was lowered from $O(l^D)$ to $O(l^{D-1})$, where $l_D$ is the volume of an effectively $D$-dimensional ACR ($(m-D)$ energy levels below the high energy ambient space of effectively $m$ dimensions). Detailed explanation of the data structure can be found in \ref{sec:step2appendix}. 

\subsection{Complexity Analysis}
\label{sec:complexity}

An formal analysis of UC variants' superior time and space complexity is provided in \ref{sec:complexityappendix}. Here we insert a comparison table indicating time and space complexity between the UC variants and Monte Carlo. Recall that $D$, the energy level or effective dimension of an ACR being sampled, is much smaller than $m$, the highest energy level or effective dimension of the ambient space where no constraints of type $C_2$ in Section \ref{sec:preliminaries} are satisfied.
 Note that the time and space complexity   is the same for all UC variants.  
 
 Since  we deal with a sampling algorithm,  it is appropriate that our complexity bounds are given in the  number of (essential) output samples,  rather than the size of the input.  It should be noted that since unused or unnecessary samples are not part of the output, the time complexity gets worse with the number of  unnecessary samples utilized by any given algorithm.  All the UC variants attempt to minimize the number of unnecessary samples.

\begin{center}
    \begin{tabular}{|c|c|c|}
\hline
     & UC & Monte Carlo \\
     \hline
    Time complexity & $ O(l^D)(linear\  in\  output\  size) $ & $ \ge O(l^{m>>D}) $ \\
    \hline
    Space complexity & $ O(l^{D-1})(sublinear\ in\ output\ size) $ & $ O(1) $ \\
    \hline
\end{tabular}
\end{center}

\section{Experiments and Results}
 Here we provide a set of experimental results that demonstrate the new UC methodology's advantages over current mainstream methods based on MC, in multiple aspects including volume computation accuracy, coverage, and efficiency of both.  See Figure \ref{fig:flowchart} in \ref{sec:guide}.
 
 For completeness, we also include comparison with ``vanilla'' EASAL, described in Section \ref{sec:previous} named EASAL-Cayley, which uses only Cayley coordinates. 
 
 We additionally compare a number of variants described in Section \ref{sec:step3}, all based on the new UC methodology's combination Cartesian and Cayley parametrization, but differing in the manner in which intersections are calculated and search directions are determined while using the frontier hypercube data structure: EASAL-hybrid-UC (the best performer all around), EASAL-UC (trades off accuracy by reducing the number of samples) EASAL-basis-UC (extremely fast and economical in the number of samples, but significantly less accurate), and EASAL-thick-UC (only for regions in which a single pair satisfies the constraint $C_2$). 

We reiterate that all EASAL-based methods use the same method for roadmapping basins. Their difference lies in whether they combine Cartesian with Cayley coordinates (EASAL-Cayley does not), and how the combination methods, collectively referred to as ``variants of UC'', perform intersections and search.

\subsection{Experiment Setup}
\label{sec:setup}
See  \ref{sec:guide} for software and data access and flowchart in Figure \ref{fig:flowchart} for reproducing the results of this paper.

The software for the algorithmic variants of UC described in the previous section were implemented atop existing curated opensource suite EASAL (Efficient Atlasing and Search of Assembly Landscapes). Software implementation of UC is available at \url{http://bitbucket.org/geoplexity/easal_dev} in branch ``feature/UniformCartesian.'' See also video \url{https://cise.ufl.edu/~sitharam/EASALvideo.mpeg}, and user guide \url{https://bitbucket.org/geoplexity/easal/src/master/CompleteUserGuide.pdf} for ``vanilla'' EASAL. 
Though EASAL is suited to full-fledged parallel processing, the experiments presented here are merely for proof-of-concept and were run on a single AMD EPYC 75F3 ``Milan'' CPU node of a supercomputer system with 40 GB of memory assigned, but all algorithmic variants of UC use less than 5GB throughout the sampling process.  Since UC variants are deterministic algorithms, a single run is sufficient as all runs produce the same output (unlike stochastic methods such as MC where a number of trajectories are required to approach ergodicity which is not guaranteed).

Comparator MC data is taken from \cite{Ozkan2021JCTC}.
We used the same set of two trans-membrane helices and Metropolis Monte Carlo trajectories from \cite{Ozkan2021JCTC} as comparator data. 
The data is available at the repository \url{https://bitbucket.org/geoplexity/easal_large_files/src/master/}. See also \ref{sec:guide}.
The model was constructed from a fragment of 11 residues and 20 atoms of the GpA homodimer structure (Protein Data Bank (PDB) ID 1F0L) with residues 73 through 83, which can be abstracted into a pair $S$, consisting of $\lvert S \rvert = k = 2$ rigid point sets $A$ and $B$, both of which have 20 points (see Figure \ref{fig:composite}). Thus the ambient dimension or highest energy level is $m = 6$. Our distance constraint systems are defined by first assigning every point $p$ in $A$ and $B$ a ``radius'' $r_p$. For all test cases the distance interval lower bound $l(a,b)$ is specified to be $0.75(r_a+r_b)$, and the distance interval upper bound $h(a,b)$ in $C_2$ is $(r_a+r_b+0.9)$. Additionally, angular restraint of the inter-principal-axis is set to be between 0\textdegree and 30\textdegree. All the constraints match those used in \cite{Ozkan2021JCTC}.  100 trajectories of about 1 million steps were used to  approach complete coverage of the grid.

\subsubsection{Generation of Baseline and Comparator  Data}

See  \ref{sec:guide} and flowchart in Figure \ref{fig:flowchart} for regenerating  the data for baseline and  UC and accessing data for MC.
We use ultra-fine uniform grid hypercubes in Cartesian space (similar to \cite{Zhang2022}) as baseline for both volume calculation and sample coverage. The grid points defining \textit{baseline grid} hypercubes are configurations satisfying the above mentioned constraints, where the 6 Cartesian translational and rotational entries of the isometry (translation and rotation matrix) are uniformly spaced at translational steps of 1\r{A} and rotational steps of $\pi/18$.

We chose Metropolis Monte Carlo (MC) and ``vanilla'' EASAL (referred to as EASAL-Cayley in Section \ref{sec:results} for clarification) as comparator methods. One rationale for the latter choice is that comparisons demonstrating this method's performance advantages over Monte Carlo/grid have already been tabulated in \cite{Ozkan2021JCTC}. However, its performance on volume computation has not been tested. Specifically, EASAL-Cayley performs the direct approach (*) described in Section \ref{sec:previous}. 

MC samples the entire landscape satisfying $C_1$.
 Since MC does not extract a roadmap of ACRs, in order to ensure fair comparisons against UC, we partition MC's samples into a \emph{pseudo-atlas} of  ACRs  using the EASAL roadmapping method. Specifically, 
we process MC sample configurations in 3 different variants (denoted as \textbf{MC1, MC2, and MC3}) to give MC the maximum advantage. 
 We partition the distance intervals in $C_2$ into two disjoint intervals \textit{inner} and \textit{outer}, with the upper bounds for inner interval set at $0.85(r_a+r_b)$, $r_a+r_b$, and $r_a+r_b+0.8$ respectively for MC1, MC2, and MC3. Each sample configuration $(A,T(B))$ satisfying $C_1$ and $C_2$ is then assigned to the ACR where $Q$ consists of atom pairs $(a \in A,b \in T(B))$ satisfying the \textit{inner} interval constraint, if $\lvert Q \rvert$ is between 1 and 6. Configurations for which $\lvert Q \rvert >6$, are assigned to the ACR where $Q' \subseteq Q$ consists of 6 pairs $(a,b)$ separated by the smallest distances. Configurations for which $Q$ is empty are assigned to the ACR where $Q'$ consists of a single pair $(a,b)$ configuration with the lowest pairwise distance. Intuitively, MC1 leans towards categorizing more configurations to higher energy level (higher dimensional) ACRs, MC3 leans opposite, and MC2 takes the middle ground.

Each experiment compares a relevant subset of the variants of UC (EASAL-UC, EASAL-hybrid-UC, EASAL-basis-UC, EASAL-thick-UC), as described in Section \ref{sec:step3}, with comparator methods w.r.t. their performance on Problems 1 and 2 of Section \ref{sec:problem_description} for basins as well as individual ACRs. 
The Cartesian hypercubes (translational and rotational step sizes $\epsilon$ in Problem 1), used in variants of UC are set to 2\r{A} and $\pi/9$ respectively. EASAL-Cayley's step size (in Cayley coordinates which are customized depending on ACR, see Section \ref{sec:previous}) is set to 0.5\r{A}.

\subsubsection{Experiment 1: Energy Basin weighted Volume}
See Figures \ref{fig:basin_volume} and \ref{fig:basin_shape}.
We chose 10 distinct energy basins for weighted volume computation accuracy and sampling efficiency comparisons (see Section \ref{sec:measurements} for measurement definitions) between EASAL-hybrid-UC, EASAL-Cayley, and the 3 variants of MC. The only criterion used in selecting these basins was that their baseline volumes were not too small.
Each \textbf{energy basin bottom} is the ACR corresponding to a set $Q$ of of $6$ constraints from $C_2$, i.e. 6 pairs in their respective energy wells. This ACR has the lowest energy level or effective dimension of 0, which is 6 levels below the ambient highest energy level, and  it consists of up to 8 rigid configurations (if any exist, see \ref{sec:previousappendix}). The basin is its upward closed set, i.e. all ACRs corresponding to all subsets $Q' \subseteq Q$ (see Section \ref{sec:preliminaries}). 

\subsubsection{Experiment 2: Individual ACR Volume and Coverage}
\label{sec:individualacr}
See Figures \ref{fig:single_region_volume}, \ref{fig:timeefficiency}, \ref{fig:coverage_accuracy}, and \ref{fig:coveragehistogram}.
We perform 4 versions of this experiment, each with 10 ACRs with $\lvert Q \rvert$ being 1,2,3,4, i.e. ACRs of dimension (energy level) 5,4,3,2 respectively. 
These ACRs were chosen from the union of the 10 basins in the previous experiment.
 All variants of MC and UC were compared for $\lvert Q \rvert = 1,2,3,4$ while EASAL-thick-UC is was used only for $\lvert Q \rvert = 1$.
 The comparisons considered volume computation accuracy, sampling efficiency, coverage accuracy and coverage efficiency; these measurements are described in detail below in Section \ref{sec:measurements}.

\subsection{Key Measurements}
\label{sec:measurements}
See \ref{sec:guide} and flowchart in Figure \ref{fig:flowchart} for  software and data (regeneration or access), and for computing the measurements and plots  of this paper.
We chose key measurements towards answering the following questions:
\begin{itemize}
    \item How do the UC variants compare to mainstream methods such as MC?
    \item Which UC variant is the best overall?
    \item Are the UC variants an improvement over  the ``vanilla''  EASAL-Cayley implementation for configurational volume computations?
\end{itemize}
All measurements listed below for each method were compared against the same measurements for the baseline.

\medskip\noindent{\bf Remark:}
It is important to note that  the fine baseline and UC  are completely deterministic methods.  
All UC variants require a single run as all runs produce the same result. MC on the other hand, being stochastic, requires several tranjectories from different initial configurations to even approach ergodicity.   To provide advantage
to MC, we took the MC dataset consisting of the union of 100 trajectories of about a million configurations each.

\subsubsection{Measurement 1a: Relative Weighted Volume of Energy Basins}
\label{sec:basinvolume}
See Figure \ref{fig:basin_volume}. 
 Weighted volume of a basin ( See Section \ref{sec:previous}) is computed by first taking the upward closed set of the basin bottom to find the ACRs in the basin. The Boltzmann factor $B$ (see Section \ref{sec:previous} for basin relative volume computation by UC and EASAL-Cayley is obtained in a manner that is consistent with MC. Specifically, we use ratio of MC's sample repetition rate between ACRs of energy level/dimension $d$ vs. $(d-1)$ (higher weight for ACRs of lower energy level or effective dimension). Accordingly, MC's Boltzmann factor was measured to be 1.068. 
 UC variants take each such ACR and compute its volume is calculated by counting Cartesian hypercubes with at least one configuration in the ACR. 
 MC sampled configurations are partitioned into the ACRs using the 3 variants MC1, MC2, MC3 as described above and the number of samples - including repetitions - is summed across all the ACRs in the basin.
  EASAL-Cayley uses the direct method (*) mentioned in Section \ref{sec:previous}, samples the convex base space in Cayley coordinates computes the pre-image Cartesian configurations, and counts them to give a rough Cartesian volume approximation. 

  The ratio of these volumes (for each method) to both the \textbf{sum} of the volumes of the 10 basins and the volume of the \textbf{union} of the 10 basins (recall that basins can nontrivially intersect, see Section \ref{sec:previous}) gives two variants of relative weighted volume.

\subsubsection{Measurement 1b: Weighted Volume Distribution across Energy Levels within Basin}
\label{sec:shapemeasure}
See Figure \ref{fig:basin_shape}. 
 An accurate volume computation  should  also recover the ``shape'' of  the basin  i.e. distribution of volume components across each energy level of the basin (summing over weighted volumes of the basin's ACRs at each energy level which decreases with number of constraints satisfied in $Q$ and increases with effective dimension $m-|Q|$). Here  we picked 1 energy basin, and for each of the methods, 4 different measures of the shape of basin were calculated. All level-specific-measurements for a given method were normalized by the basin's weighted volume as calculated by the same method. The 4 shape measures are defined below. Here $B$ is the Boltzmann factor obtained as described above.
\begin{enumerate}
    \item Weighted volume at each energy level.   Denote by $M(k)$  the number of samples of method $M$ in energy level or effective dimension $k$.   Plot $M(k)B^{5-k}/\Sigma_j M(j)B^{5-j}$  against the levels $k$.
    \item Weighted volume at each energy level relative to weighted volume of all ACRs at that level in the union of basins.  
    Denote by $rv_M(k)$ the number of samples of method $M$ at dimension/energy level $k$ divided by total number of $M$'s samples at  energy level $k$ over the disjoint union of  all ACRs in all the 10 basins at the same energy level $k$.  Plot $rv_M(k)B^{5-k}/\Sigma_j rv_M(j)B^{5-j}$ against $k$.
    \item Weighted volume at each energy level relative to the average ACR at that level.  Here we use the average of  over  ACR's at   energy level $k$ (instead of number of total samples on the level as used in the previous item), therefore the distribution is in effect computing the number of typical or average  ACRs of each given  energy level across all  energy levels. Denote by $rva_m(k)$ the number of method $M$'s   samples across all ACRs at energy level $k$ in the basin divided by the average number of method $M$'s  samples in all the  ACRs at that level across all basins. Plot $rva_M(k)B^{5-k}/\Sigma_j rva_M(j)B^{5-j}$ against $k$.
    \item Weighted volume at each energy level relative to the  basin average volume at that level. (Using  basin averages (instead of ACR averages  used in the previous item) in computing relative volumes).   Denote by $rvab_M(k)$ the number of method $M$'s   samples at energy level $k$ for this basin divided by average number of $M$'s   samples over all basins at energy level $k$. Plot $rvab_M(k)B^{5-k}/\Sigma_j rvab_M(j)B^{5-j}$ against $k$.

\end{enumerate}

\subsubsection{Measurement 2a: Individual ACR Volume  Accuracy}
\label{sec:single_region_volume}
See Figure \ref{fig:single_region_volume}. 
 As described in Section \ref{sec:individualacr} we compute the ratio of the volume of  each ACR   - in each of the 4 groups - w.r.t. the total volume of the respective group.  

\subsubsection{Measurement 2b: Sampling Efficiency (for basins and ACRs)}
\label{sec:single_region_efficiency}
See Figures \ref{fig:basin_volume} and \ref{fig:timeefficiency}. 
We plot the number of  samples  required by each method for Measurements 1a and 2a. 
We additionally measure actual time per sample used in Measurement 2a for UC and EASAL.  The latter measurement is not available for MC since experiments of EASAL and UC and those of MC  were executed on different computational platforms.

\subsubsection{Measurement 2c: Coverage Accuracy and Efficiency (for ACRs)}
\label{sec:single_region_coverage}
See Figures \ref{fig:coverage_accuracy} and \ref{fig:coveragehistogram}. 
Coverage by a method $M$ of the baseline Cartesian grid is measured using $\gamma_M$-coverage  in order to normalize methods with different number of samples. A $\gamma_M$-hypercube is a hypercube with a baseline grid point as center and $2\gamma_M$ as range for each Cartesian coordinate. A baseline grid point $p$ is \textit{covered} by method $M$ if at least one sample point (center of feasible hypercube for UC) found by $M$ lies within  a $\gamma_M$-hypercube centered around $p$. We define $\gamma_M$-coverage  as the percentage of baseline grid points covered, and the value of $\gamma_M$ is set to
\begin{center}
$\gamma_M:= (\frac{\Gamma}{\sigma_M})^{1/6}$
\end{center}
where $\Gamma$ is baseline grid point count and $\sigma_M$ is sample point count for Method $M$.

Two aspects of coverage, accuracy (see Figure \ref{fig:coverage_accuracy}) and efficiency (see Figure \ref{fig:coveragehistogram}), are measured in the experiments. Accuracy (in fact, error) is measured by counting the number of baseline grid points \textbf{missing} in the coverage, i.e. no sample point lies within its $\gamma$-hypercube. Efficiency of coverage is measured by number of sample points in each $\gamma$-hypercube. Intuitively, an efficient method would only  sample 1 point in each $\gamma$-hypercube.

\subsection{Results}
\label{sec:results}
See \ref{sec:guide} for instructions on reproducing results in this paper, including software for UC data generation, MC data repository,  scripts for processing data for experimental measurements listed below, and  generating plots.

The results of all experiments and measurements are listed below. Overall, EASAL-hybrid-UC and in fact all variants of UC perform significantly better than both MC and EASAL-Cayley  for the experiments and measurements of this paper, related to volume or configurational entropy.

\subsubsection{Energy Basin Volume}

\begin{figure}[htbp]
\centering
    \includegraphics[width=\linewidth]{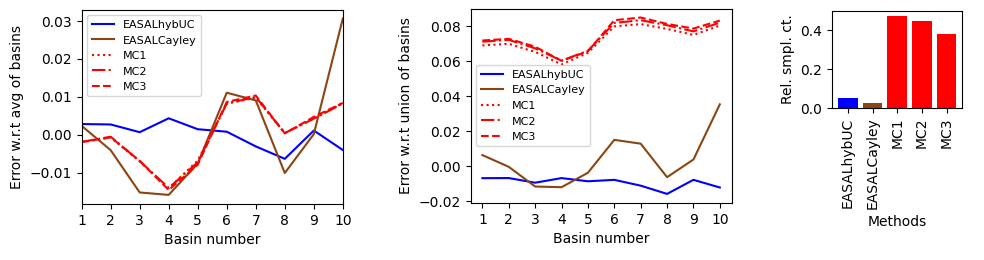}
\caption{Volume result for basins}
\scriptsize Relative volume w.r.t average of basins (left) and union of basins (mid), deviation from baseline, closer to 0 means better. Number of samples for each methods on right. See Section \ref{sec:basinvolume} for details.
\label{fig:basin_volume}
\end{figure}
As described in Section \ref{sec:basinvolume}, the plots in Figure \ref{fig:basin_volume} show that all methods  compute  the volume ratio between   energy basins reasonably well. However, when the denominator switches to the union of 10 basins, MC (regardless of variants) reports values much higher than baseline. This is  because MC loses precision within basins as it does not compute a roadmap of the basin. As a result, for the 10 basins of this experiment,  MC overestimates  the volume of ACRs  shared by the basins (see Section \ref{sec:previous}),  which have higher energy level,  effective dimension and volume,  while  underestimating  the weighted volume of the ACRs unique to the basin, which are closer to the basin bottom and have a lower energy level, effective dimension and volume. It is also worth noting that EASAL-Cayley also shows a ``fa\c{c}ade'' of acceptable results, mainly because of the law of large numbers:   basins being a large collection of  ACRs,   overestimation and underestimation errors cancel each other out to some degree, making the final result appear miraculously  good. We will see that this luck does not hold in volume experiments for single ACRs. 

\subsubsection{Energy Basin Volume Distribution}

\begin{figure}[htbp]
\centering
    \includegraphics[width=\linewidth]{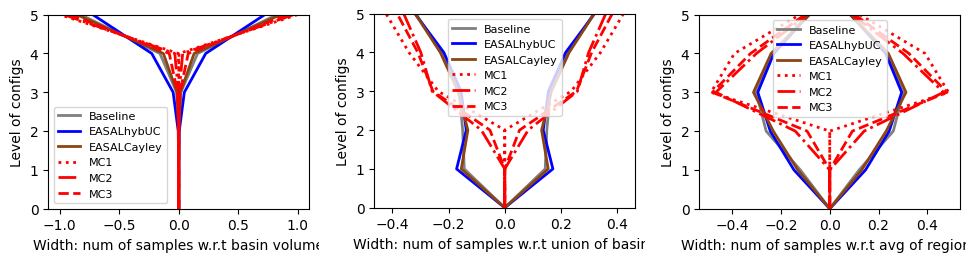}
\caption{Weighted volume distribution across energy levels or effective dimensions for one selected basin}
\scriptsize Wider means a  larger portion of weighted volume concentrated at that level; volume on each level (left), relative to union of basins (mid), relative to average of regions (right) on each energy level. Gray: baseline, blue: EASAL-hybrid-UC, brown: EASAL-Cayley, red: variants of MC. Closer to baseline is better. See Section \ref{sec:shapemeasure} for detailed description.
\label{fig:basin_shape}
\end{figure}

\begin{figure}[htbp]
    \centering
    \includegraphics[width=.66\linewidth]{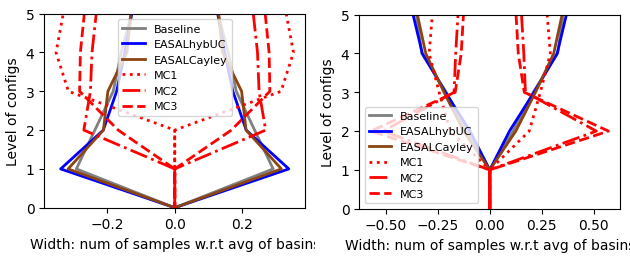}
    \caption{Weighted volume distribution across energy levels or effective dimensions, comparison for two basins}
    \scriptsize Volume distribution relative to average of basins for two different basins. Note that left is the same basin shown in Figure \ref{fig:basin_shape}. See Section \ref{sec:shapemeasure}.
    \label{fig:basin_shape_alt}
\end{figure}

The results for Measurement 1b in Section \ref{sec:shapemeasure} are shown in Figure \ref{fig:basin_shape}. MC has difficulty  estimating volumes of lower energy or effectively lower-dimensional ACRs, which is clearly illustrated in these images showing MC's concentration of weighted volume proportional to energy level or effective dimension. Specifically for distribution relative to average over basins, an ``average'' or ``typical'' basin should have a uniform flat distribution, i.e. all entries at 20\%. The baseline shows that basin 1 in Figure \ref{fig:basin_shape} has more weighted volume concentrated on lower energy   ACRs of effectively lower dimension, while all 3 variants of MC report the opposite; and to further illustrate MC's poor performance, we show another basin (basin 3) in Figure \ref{fig:basin_shape_alt}, where the baseline shows that the weighed volume distribution across levels is reversed, but again MC reports the opposite. In contrast, despite its relatively low resolution, due to the Cayley parametrization that underpins the sampling, UC finds and directly samples lower energy ACRs of effectively lower dimension  in their intrinsic dimension accurately, but mildly underestimates volume at higher energy levels. 

\subsubsection{Single Region Volume and Efficiency}
\begin{figure}[htbp]
\centering
    \includegraphics[width=.8\textwidth]{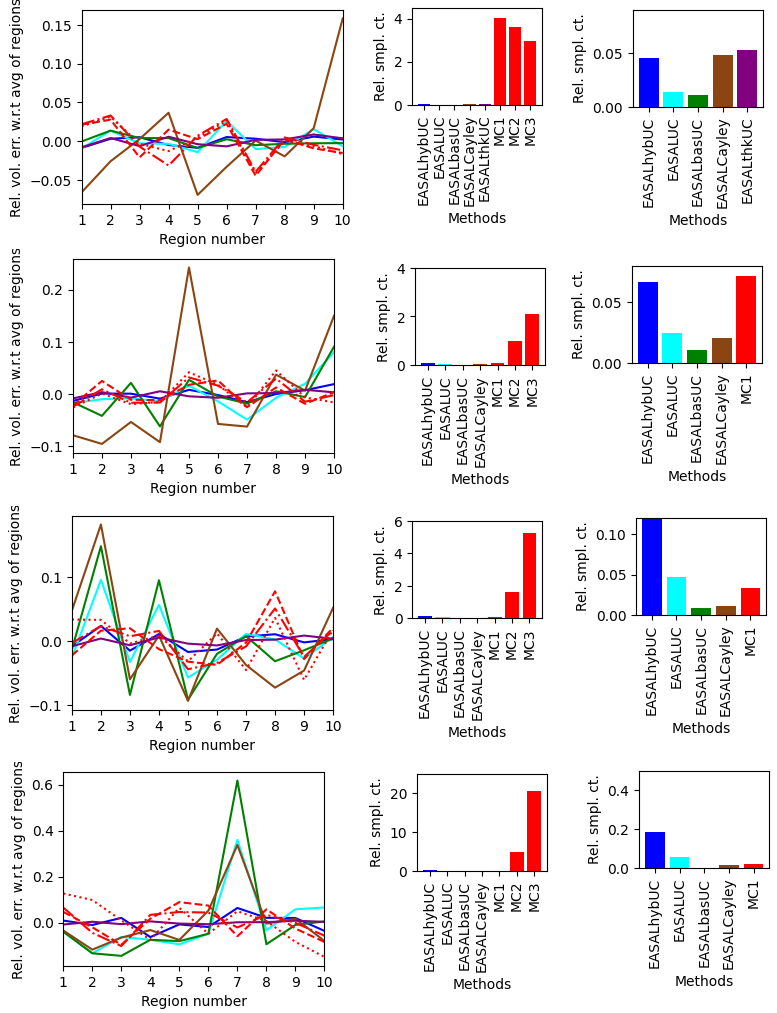}
\caption{Volume accuracy and efficiency for individual ACRs}
\scriptsize ACRs at energy level / effective dimension 5  (1st), 4  (2nd), 3  (3rd), and 2 (last row): Left: relative volume of EASAL-hybrid-UC (blue), EASAL-UC (cyan), EASAL-basis-UC (green), EASAL-Cayley (brown), EASAL-thick-UC (purple, energy level 5  only), MC (red) w.r.t the average of regions, deviation from baseline. Closer to 0 means better. Mid: relative sample count w.r.t the baseline. Right: relative sample count of UC variants magnified 50 times.
See Sections \ref{sec:individualacr} and \ref{sec:single_region_volume} for details.
\label{fig:single_region_volume}
\end{figure}

Results for Measurement 2a, as described in Section \ref{sec:single_region_volume}, are shown in Figure \ref{fig:single_region_volume}. EASAL-hybrid-UC performs better than all 3 variants of MC by a steady margin. All UC variants' volume computation is much more accurate than EASAL-Cayley. This shows that Cayley sampling is necessary, but insufficient, and the additional ingredients in UC are necessary. EASAL-Cayley's deviations from the baseline cancel each other across the different ACRs which indicates that for aggregate basin volume computations, it may not be a bad choice. This is consistent with the fact that this direct ``vanilla'' EASAL method of approximating basin volumes using mere Cayley sampling (without considering Cartesian volumes at all) provided predictions \cite{Wu2020PLoS} of crucial interactions for virus capsid assembly which were experimentally validated. 

Among variants of UC, EASAL-hybrid-UC shows the best volume accuracy. This is mainly because of its ability to greatly mitigate error in linearization. In some cases, especially in lower energy ACRs of lower effective dimension, both EASAL-UC and EASAL-basis-UC miss a large portion of the total volume, corrected by EASAL-hybrid-UC. It is worth noting that all variants of UC deviate in a correlated manner for certain ACRs, both over/under-estimating. We speculate this is caused by those ACRs being narrow in one (or more) of the Cayley or Cartesian coordinate directions, thus they are closer to effectively even lower-dimensional entities. This magnifies the error due to relatively coarse resolution we use.

EASAL-hybrid-UC is overall the best method as it performs better than all 3 variants of MC by a steady margin, and uses far fewer samples than MC. 

Figure \ref{fig:single_region_volume}'s right 2 columns show number of samples utilized by each method over all the 10 test cases for each energy level or effective dimensionality of ACRs. See Section \ref{sec:individualacr}. 
It is clear that UC reaches a high level of volume accuracy with much fewer samples than MC. Also worth mentioning is that in lower energy or effectively 3 and 2 dimensional ACRs, MC1 ( which uses  the ``thinnest'' upper limit for partitioning samples) shows a plunge in the number of samples from energy levels/dimensions 5 to 3 , mainly because it lacks the ability to directly sample the narrow, lower, but constant energy, effectively lower-dimensional ACR and has to often visit the higher energy 6 dimensional ambient space around it  to rule out those  configurations. Although MC1 appears to do well in the basin aggregate in Figure \ref{fig:single_region_volume}, it is clear from the anomalous plunge in sample points from energy level/dimension 5 to 3 to 2, as well as the basin plot in Figure \ref{fig:basin_shape}, and the coverage error plot (Figure \ref{fig:coverage_accuracy}) that this apparent performance in Figure \ref{fig:single_region_volume} is misleading. This strengthens our posited  advantage of EASAL based methods' over MC: MC sampling cannot remain entirely within a narrow, lower constant energy, lower-dimensional  ACRs, without sometimes visiting adjacent somewhat higher dimensional  ACRs, hence most of the samples used to calculate the lower dimensional  ACR's volume do not actually lie in the  ACR, but close to it.  Close to the limiting cases when the Lennard Jones well   is narrow, the  ACR could be missed completely; and although setting a wider interval for the well could help the sampling,   tracking the low constant energy or low dimensional region  no longer happens  which makes such a method prone to errors.

\begin{figure}[htbp]
\centering
    \includegraphics[width=\linewidth]{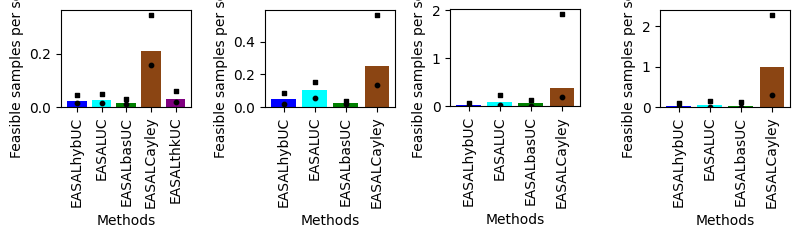}
\caption{Time efficiency for single regions}
\scriptsize
Useful samples (that contribute to volume) found per millisecond for ACRs at 5 (1st), 4 (2nd), 3 (3rd), and 2 (4th)-energy level (effective dimension) using each method: EASAL-hybrid-UC(blue), EASAL-UC(cyan), EASAL-basis-UC(green), EASAL-Cayley(brown), EASAL-thick-UC(purple, 5th energy level only). Higher means better. See Section \ref{sec:single_region_efficiency}.
\label{fig:timeefficiency}
\end{figure}

Figure \ref{fig:timeefficiency} shows results for Measurement 2b in Section \ref{sec:single_region_efficiency} with time cost per useful sample point, i.e. that contributes to volume calculation for UC variants and EASAL-Cayley. As mentioned, MC is not included as the results were obtained on a different computational platform. The result shows that UC's superior volume calculation accuracy comes at a significant expense w.r.t. time, although the formal complexity analysis indicates no difference, i.e. linear time complexity in the output size for all methods (see Section \ref{sec:complexity}). 

Among UC variants, EASAL-basis-UC handles each sample quicker when energy level or dimensionality gets lower. Because of its coarser hypercube decomposition procedure, using a hyper-parallelepiped (rather than simplex) level decomposition, the number of linear system solving calls are significantly lowered, which is further augmented when the ACR's energy level and dimensionality is low hence decomposing hyper-parallelepiped into simplices takes significantly more time, which basis-UC completely skips. 

 On the other hand, EASAL-hybrid-UC pays in  time spent to fix the linearization error. Results on time-per-point ratio show the accuracy-efficiency trade-off: EASAL-hybrid-UC is more accurate (while spending significantly more time to generate each sample point) and EASAL-UC is faster, giving potential users another parameter to customize based on their needs.

\subsubsection{Coverage Error and Efficiency}

\begin{figure}[htbp]
\centering
    \includegraphics[width=\linewidth]{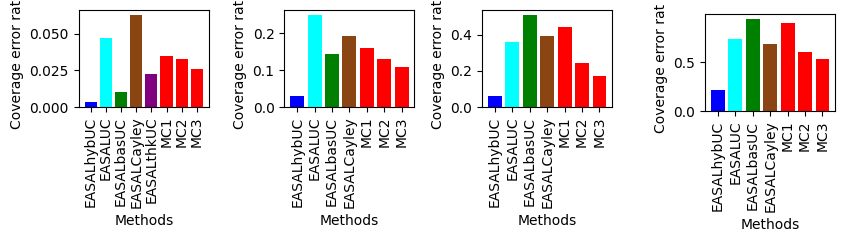}
    \caption{Coverage error percentage}
    \scriptsize Ratio of baseline points missed for 5 (1st), 4 (2nd), 3 (3rd), and 2 (4th)-dim configuration spaces using each method: EASAL-hybrid-UC (blue), EASAL-UC (cyan), EASAL-basis-UC (green), EASAL-Cayley (brown), EASAL-thick-UC (purple, 5-dim only), MC (red). Lower means better. See Section \ref{sec:single_region_coverage} for details.
    \label{fig:coverage_accuracy}
\end{figure}
As shown in Figure \ref{fig:coverage_accuracy}, EASAL-hybrid-UC shows significantly better coverage for the sampled ACR in comparison to other methods across the board. Specifically for lower constant energy or lower dimensional regions, it is the only reliable method fulfilling the task of covering such regions with a set of sample points, for which traditional methods such as MC struggle. 

\begin{figure}[htbp]
\centering
    \includegraphics[width=\textwidth]{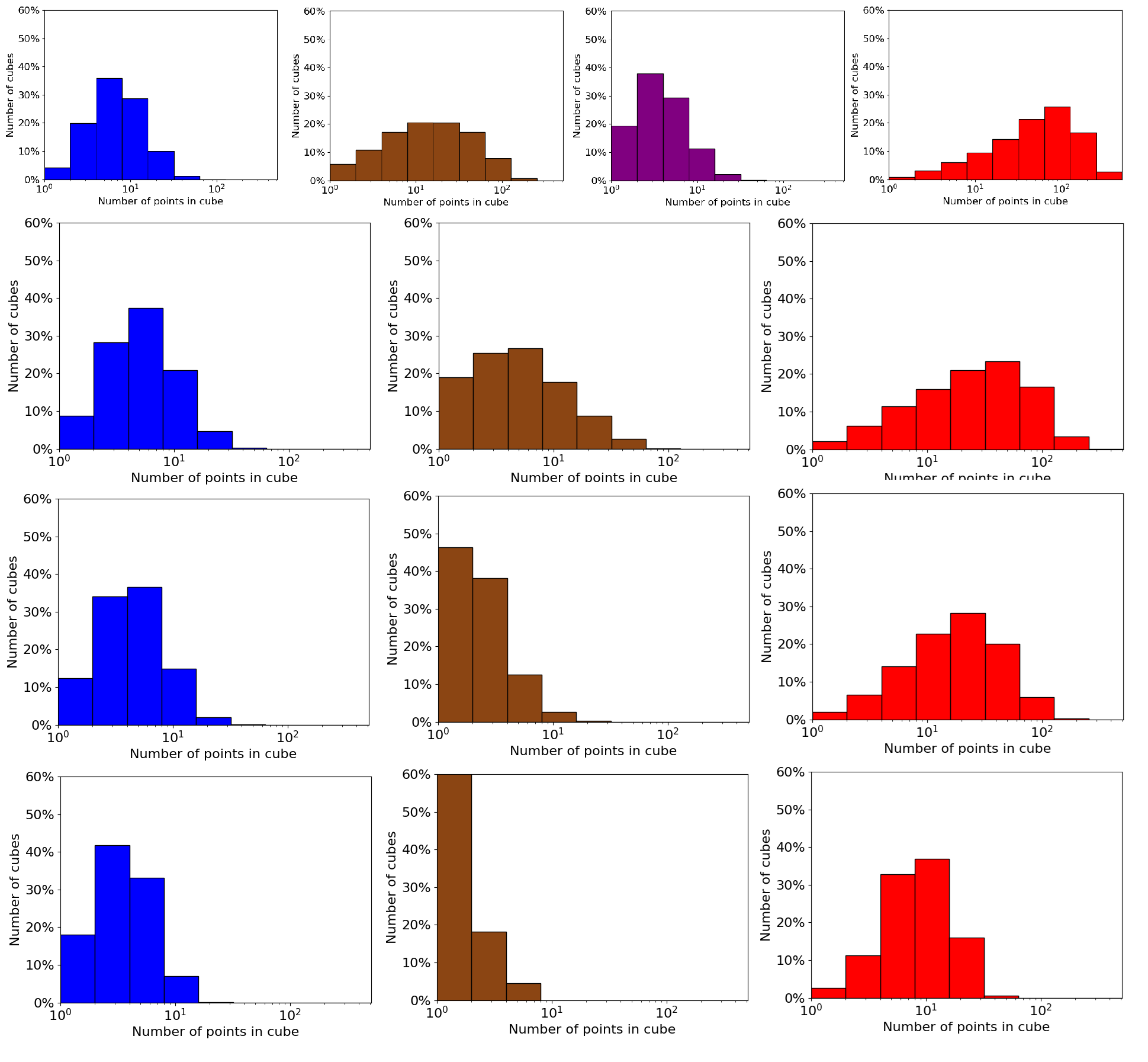}
\caption{Coverage Efficiency}
\scriptsize Coverage efficiency for test case ACRs at energy levels 5 (1st), 4 (2nd), 3 (3rd), and 2 (4th row)   using each method: EASAL-hybrid-UC (blue), EASAL-thick-UC (purple, 5-dim only), EASAL-Cayley (brown), MC2 (red). In each subplot, horizontal axis shows the number of sample points $\nu$ that lie in $\gamma$-cubes and vertical axis shows the fraction of $\gamma$-cubes having $\nu$ mapped points in them.  More efficient methods have higher bars to the left and lower to the right. See Section \ref{sec:single_region_coverage} for details.
\label{fig:coveragehistogram}
\end{figure}

For Measurement 2c (in Section \ref{sec:single_region_coverage}), the results are shown here in Figure \ref{fig:coveragehistogram}. We plot the number of sample points $\mu$ that lie in an $\gamma$-grid-hypercube around a baseline grid point, against number of $\gamma$-hypercubes with $\mu$ sampled points. A more efficient method should have fewer points mapped to the same $\gamma$-cube, i.e. higher bars on the left side of the plot. As is seen in the plot, both EASAL-hybrid-UC and EASAL-Cayley show reasonably good results thanks to them directly sampling the constant lower energy, lower effective dimensional region. However, EASAL-Cayley's good performance also comes with the fact that it is missing a large portion of some sampled ACRs (see Figure \ref{fig:coverage_accuracy}). For the sake of simplicity, we picked MC2 for this plot being the overall balanced performer between sample size and accuracy. Its performance is not as good as EASAL-hybrid-UC (due to its large number of samples), and it also comes with lower accuracy when energy level or dimension of ACR drops. This strengthens our claim that UC, especially EASAL-hybrid-UC, is superior to MC for the configurational entropy computation related tasks studied here.

\section{Discussion and Future Work}
Overall, EASAL-hybrid-UC and in fact all variants of UC perform significantly  than MC for the experiments and measurements of this paper, related to accuracy and efficiency of volume or configurational entropy computations. They also perform better than the ``vanilla'' version, EASAL-Cayley, showing the importance of using a judicious mix of Cayley and Cartesian sampling.
Next, we suggest potential improvements.

(1) Given the convex base space in Cayley coordinates, the randomized sampling for computing volumes of convex bodies given by \cite{Dyer1991, applegate1991, kannan1997, LOVASZ2006, Ge2015AFA} could potentially be used to solve Problem 1 directly, thus bypassing Problem 2. However, to translate this to an accurate computation of the Cartesian covering space poses a challenge: although the covering map is quite well behaved, this cannot be said about the pseudoinverses of its Jacobian or Hessian. 

(2) The tradeoff between accuracy and efficiency both in volume computation and coverage is clearly demonstrated in the comparisons between the different UC variants as well as between all UC variants
that use Cayley coordinates only to reduce dimension and guide sampling, and EASAL-Cayley that relies on sampling entirely in Cayley coordinates. Further hybridizing these methods in a manner appropriate to requirements of specific applications is indicated. Similar complementarities with MC and other prevailing methods should be exploited by hybridizing with those methods for specific applications.

(3) Optimization of the implementation of all UC variants is expected to reduce numerical errors and parallelization is expected to improve performance on larger number of rigid molecular components $\lvert S \rvert >> 2$ studied in this article.

(4) While the emphasis of this paper was to explain the key underlying conceptual details  of UC and provide both formal guarantees and proof-of-concept comparisons, it remains to test the UC methodology for binding affinity, and hot-spot residue computations on well known benchmark datasets for computational alanine scanning, ligand docking and Lennard-Jones clusters \cite{trombach2018, jankauskaite2018skempi, argawal2019docking}.
The longer term goal is to demonstrate use of the UC methodology's efficient computation of the above quantities to make concrete progress on specific, poorly understood, soft-matter assembly problems.

\section{Data and Software Availability}
Please see  \ref{sec:guide} for detailed instructions to access data and software for reproducing the results of this paper.


\section{Acknowledgments}
The authors  thank Aysegul Ozkan and Rahul Prabhu for many fruitful discussions on leveraging EASAL for configurational entropy and volume computations. The authors were partially sponsored by NSF DMS 1563234 and NSF DMS 1564480.

\appendix

\section{Guide for reproducing results of this paper}
\label{sec:guide}
\begin{figure}[htbp]
\centering
    \includegraphics[width=.8\textwidth]{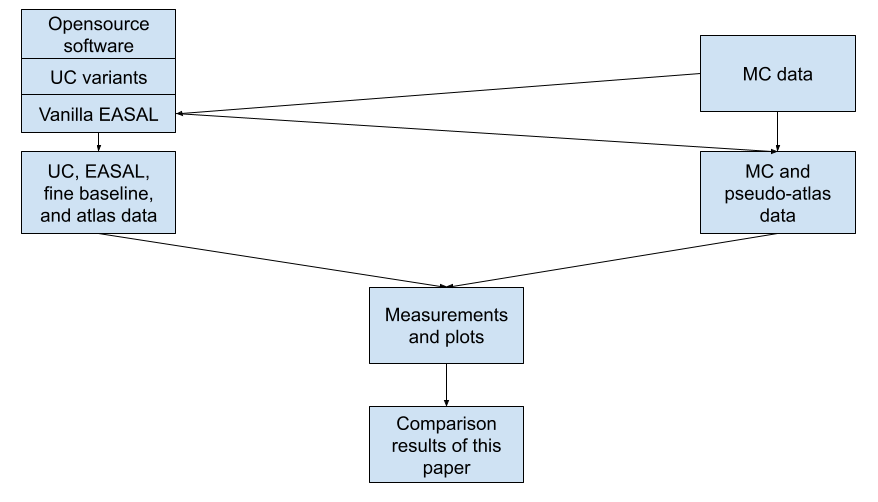}
\caption{ Overview flowchart for reproducing the results of this paper} 
\label{fig:flowchart}
\end{figure}

See flowchart in   Figure \ref{fig:flowchart} for an overview.

\subsection{Reproducing EASAL-UC data}
\begin{enumerate}
    \item Clone from EASAL software: \url{https://bitbucket.org/geoplexity/easal-dev/src} in branch \\``feature/UniformCartesian'', install EASAL and dependencies following the readme, compile using the backend version.
    \item Sample files are stored in \textit{files} folder and can be used as test cases for a demo run. Use \\ \textit{3AtomsA/B.pdb}, \textit{6AtomsA/B.pdb}, \\ \textit{initial\_test\_small\_helixes\_A/B.pdb}, and \textit{initial\_testA/B.pdb} for 3, 6, 20, and 42 points in each point set. Notice that the experiments performed in this article used \textit{initial\_test\_small\_helixes\_A/B.pdb}.
    \item In \textit{settings.ini}:
    \begin{itemize}
        \item Put test case name in \textbf{PointSetA} and \textbf{PointSetB - file}. 
        \item For a certain node (defined by a set of contacting pairs between A and B)'s volume, set \\ \textbf{RootNodeCreation - dimension\_of\_initialContactGraphs} to the dimension of such node (i.e. 6 minus number of contacts), and write each of the contacting pairs in \textbf{Sampling - initial\_Contact\_XX}.
        \item Bounds of each constrained pair $(a, b)$ is calculated using formula $d = \lambda (r_a + r_b) + \delta$. 3 sets of $(\lambda, \delta)$ pairs are defined in \textbf{Constraint - activeLower, activeUpper, collision}. Set \textbf{activeLower} and \textbf{collision} the same, and \textbf{activeUpper} larger than them, to avoid unexpected behavior for UC. 
        \item \textbf{Sampling - cartesianIntersectionMode} is used to pick from different varieties of decomposition methods:
        \begin{itemize}
            \item 0 - EASAL-UC
            \item 1 - EASAL-basis-UC
            \item 2 - EASAL-thick-UC
            \item 3 - EASAL-face-UC
            \item 4 - EASAL-hybrid-UC
        \end{itemize}
        \item Sampling resolution can be set by changing \textbf{Sampling - cartesianSteps}. The first 3 entries stand for translational step size, and the latter 3 for rotational step count per $2\pi$ (for example, step count of 4 equals angular step size of $\pi/2$, or 90\textdegree). 
    \end{itemize}
\end{enumerate}

\subsection{Generating UC measurements and plots}
Python scripts used for generating the  measurements and plots using the data generated by the UC software are provided in the "plot" folder of the repository. 

\subsection{Reproducing MC data}
Raw MC data files are available at \url{https://uflorida-my.sharepoint.com/:f:/r/personal/yichi_zhang_ufl_edu/Documents/easal_mc_log?csf=1&web=1&e=L3m65d} (as these are large trajectory data files, please email authors to request permission to download). Each line in the file is a sampled point in MC in the form of a 6-tuple: first 3 entries for translations and last 3 for rotations.

\subsection{Generating MC measurements and plots}
MC data files (see above) can be processed using EASAL software with the \\ \textbf{Sampling-UniformCartesianMode} in \textit{settings.ini} set to 5, which will generate the pseudo-atlas of the MC data and output volume of each individual ACR in this pseudo-atlas (see description in \ref{sec:setup}). The result can then be processed using Python scripts provided in the "plot" folder of the repository. 

\section{Mathematical treatment of Methods Section}
Here we provide a largely self-contained,   rigorous treatment of the mathematics underlying the Methods section of the main article. 
\subsection{Preliminaries}
\label{sec:preliminariesappendix}
We start with the following precise formulation of assembly using distance constraints in order to formalize and solve the problem of volume computation of configurational regions.

An \textit{assembly constraint system} $C$ is specified by:
\begin{itemize}
    \item A finite set $S = \{S_1, S_2, ..., S_k\}$ of $k$ point sets in $\mathbb{R}^3$.   
    \item A set of distance (interval) constraints between points in different $S_i$.
\end{itemize}
The variables are the \textit{Euclidean/Cartesian} isometries $T_{S_i}$ for each $S_i \in S$, given by  6 scalars specifying $S_i$'s rotation and translation relative to some fixed $O \in S$, where $T_O$ is assumed to be the identity.
 
For our intended application, each of the constraints belongs to either of the following categories:
\begin{itemize}
\item $C_1$: $\forall A \in S, B \in S, A \neq B, \forall (a, b), a \in A, b \in B$, $\lVert T_A(a)-T_B(b) \rVert \ge l(a,b)$. That is, every pair of points belonging to different point sets has a lower bound on the distance between them. Intuitively, these constraints prevent collision between solid particles (atoms, residues, etc). 
\item $C_2$: $\exists (a,b)$ where $a \in A, b \in B, (A, B)\in S$, $\lVert T_{A}(a)-T_{B}(b) \rVert \le h(a,b)$. In other words, some pairs are constrained by distance upper bounds, and combined with $C_1$, by distance intervals. Constraints in this group prohibit the point sets being arbitrarily far away from each other where their interaction would cease.
\end{itemize}
The \textit{configuration space}, denoted $R_S$, is the set of configurations $\{T_{S_i}\}$ such that all constraints are satisfied. Our goal in this article is to efficiently and accurately calculate the (relative) volume of \textit{configurational regions} or subsets of $R_S$.  
For the problems considered in this paper, we will assume -- essentially without loss of generality -- the limiting case where the interval size $(h(a,b) - l(a,b))$ tends to 0, as this case is both more difficult and more interesting; and (ii) generically configuration space $R_S$ is non-empty, infinite, and bounded \cite{graver1993combinatorial, sitharam2018handbook}. Furthermore, while this paper deals only with Euclidean distance constraints, the concepts generalize to non-Euclidean norm or even other metric distance constraints. 

Typically, Cartesian configuration spaces for systems mentioned above are semi-algebraic sets in high dimensional ambient space ($k$ point sets in  Euclidean dimension 3 give $m = 6(k-1)$ ambient dimensions for configuration space $R_S$), and generically, when the constraints in $C_2$ are ``narrow'', e.g. $(h - l)$ is close to 0, the configuration space is a real algebraic (quadratic) variety of co-dimension 1 (dimension one less than the ambient space)\cite{graver1993combinatorial, sitharam2018handbook}. 
This is because when interval constraints are narrow, each of them being satisfied effectively reduces the dimension of the configuration space by one. So that we are dealing with a configuration space that is effectively of much smaller dimension than the ambient dimension. For example $(m-1)$ (resp. $(m-2)$) such ``small'' interval constraints would generically yield a configuration space that is a  curve (resp. sheet) in the ambient $m$-dimensional space, with some ``thickness'' which tends to 0 in the limiting case. See examples in Figure 3 of the main article.
This assumption holds in our intended applications, typically, since the distance constraints are either lower bounds, or, if they are bidirectional then the specified intervals of allowable distances are typically ``small'' as indicated in the assumption above.

Traditionally, solving such a system requires traversal of an $(m-1)$-dimensional, topologically complex $R_S$ in Cartesian configuration space of isometries $T$. As mentioned in Related Work subsection of the main article, dealing with distance equality or inequality constraints, the task of traversing or sampling configurations while staying within the effectively lower dimensional  Cartesian configuration space is achieved in a standard setting using an onerous type of ``gradient descent'', i.e. repeated linear tangential steps (e.g. by computing the Jacobian of the distance map of the constraint system), alternating with projections/corrections back to the feasible region. When prevailing methods are used, including molecular dynamics or Monte Carlo based methods, this gradient descent process pervades many common tasks such as finding optimal or extremal configurations, finding paths, path lengths, region volumes (configurational entropy), path probabilities for transition networks, sampling configurations or paths etc. Such traversal is impractical when the number of points in the point sets and number of  distance interval constraints in $C_2$ is large. We simultaneously face the combined curses of ambient dimensionality and topological complexity, calling for a novel approach for mitigating such issues.

\subsection{Mathematical details of previous work on EASAL}
\label{sec:previousappendix}

\subsubsection{Roadmap and Basins}
The problem of satisfying the constraints in the constraint system $C$ described in Section 2.1.1 of the main article can be divided into subproblems as follows. Notice that the existential quantifier in $C_2$ can be replaced by a disjunctive union over all simultaneously satisfied subsets $Q$ of distance  upper bound (hence interval) constraints $C_2$. Consider a variant of the system consisting of all collision-avoidance constraints in $C_1$, but replacing $C_2$ by the conjunction of  distance upper bound constraints in a subset $Q$. The feasible configuration space \cite{OzkanBiCoB2011,Ozkan2018ACMTOMS, Prabhu2020JCIM} $R_S$ of the original problem is partitioned into disjunctive union of ACRs $R_{S,Q}$, which is defined as the set of all configurations $T$ satisfying \textbf{all} constraints in $C_1$ and $Q$. The set of ACRs is stratified by the size of $Q$ and organized as a partial order (in the form of a directed acyclic graph (DAG)) called the \textit{roadmap}. Each node of the roadmap DAG represents an ACR, and directed edges (or parent-child relationships) reflect the superset-subset relationship of the ACRs (and when the limiting case is reached, the relationship becomes interior-boundary), or alternatively the subset-superset relationship of the corresponding distance interval constraint sets $Q$ from $C_2$. That is, a child node’s ACR is a subregion of its parent, and it always satisfies exactly one extra distance interval constraint than its parent. Notice that since each constraint corresponds to the system being within one discretized Lennard-Jones potential well, each ACR in the roadmap also represents a contiguous configurational region with constant energy level. If we further assume that all the Lennard-Jones potential wells are of equal, unit depth (without loss of generality), the level of stratification of an ACR also corresponds to the relative energy for configurations in it. 

Both the DAG and stratification structures of the roadmap are used for exploration by the algorithm outlined below. This decouples the exploration of the roadmap from sampling process of each node in the roadmap.

Observe that energy basins are subsets of the roadmap, specifically \textit{upward closed sets} for given ACR with constraints $Q$ as the bottom of the basin. In other words, the basin is the set of all the ACRs with active constraint set $Q'$ with $Q' \subseteq Q$, i.e., the collection of all the roadmap ancestors of the ACR at the basin bottom. In the terms of energy, the bottom is the unique energy minimum region of the basin, and there exists a direct, barrier-less, energy-lowering path starting from any ACR in the basin down to the bottom ACR. Recall that a parent-child (or ancestor-descendant) relationship also leads to the interior-boundary relationship for the ACRs, the ACR at the basin bottom is the shared boundary of all other ACRs in the basin. 

 We note that basins are not necessarily disjoint sets and can share the same ACRs. Different views of an energy basin are shown in Figure 2 of the main article.
For free energy computation, if we assume that each satisfied constraint in $C_2$ (corresponding Lennard-Jones well) reduces energy by the same amount, then the appropriate configurational entropy measure for a basin is the weighted volume, obtained by equally weighting volumes of the ACRs  at each energy level  or effective dimension within the basin. Specifically, given a Boltzmann factor $B$ associated with each energy level, the weighted volume of a basin is $V=\sum_{k=1}^{m-1}B^{(m-1-k)} V_k$, where $V_k$ is the summation of (unweighted) volumes of all ACRs at energy level $k$  in the basin.

\subsubsection{Cayley Parametrization: Motivation}

Now we focus on each ACR $R_{S, Q}$ given a set of constraints $Q$. As stated before, the energy level, i.e. the effective dimension of $R_{S, Q}$ is potentially equal to or lower than the ambient dimension $m$, and  is topologically complex in Cartesian (see Figure 1 of the main article) arising from the combination of $C_1$ and $Q$. 

Cayley parametrization \cite{SitharamGao2010, SitharamWilloughby2015}, defined formally below, is broadly applicable \cite{SitharamWang2014Beast, Wang2015, Sadjadi2021}, and maps $R_{S,Q}$ into a convex, and hence topologically simple base space of the same dimension, consisting of Cayley configurations. In the limiting case where $(h-l)$ is close to 0 and the ACR $R_{S,Q}$ is ``thin,'' Cayley parametrization further ``flattens'' it into a $(m-\lvert Q \rvert)$ dimensional space residing in ambient space of the same dimensionality. 

 Cayley parametrization is further motivated by the following desirable properties. The combination of these properties ensures high sampling efficiency and accuracy, avoiding gradient-descent or retraction maps to enforce constraints, avoiding repeated sample configurations and minimizing discarded sample configurations.
\begin{itemize}
    \item The Cayley parametrization is a  \emph{covering} map from Cartesian ACR $R_{S,Q}$, called a \emph{branched covering space} onto the convex \emph{base} space which is computationally easy (constant time), essentially just measuring distances within a Cartesian configuration $T$ which have not been explicitly constrained (see formal definition below).
    \item The boundaries of the convex Cayley base space are easy to compute, and depend entirely on the distance interval constraints in $Q$. Additional collision avoidance constraints in $C_1$ carve out typically a small number of convex Cayley parametrizable regions from this convex Cayley base space.
    \item The inverse (from Cayley to Cartesian) maps each Cayley configuration in the base space to generically finitely many Cartesian configurations  each uniquely identifiable by chirality, and easily computable (constant time). The collection of pre-image Cartesian configurations corresponding to each chirality is called a \emph{flip}. Nongeneric  Cayley configurations could have pre-images in multiple flips or  could have infinitely many pre-image configurations, but these constitute a vanishingly small, measure zero set of Cayley configurations which can be ignored for the purposes of this paper.
\end{itemize}
Using these properties, one can efficiently traverse the lower constant energy ACR that is effectively lower dimensional and topologically complex in Cartesian  $R_{S, Q}$ living in high ambient dimension as follows. Traversing the lower dimensional convex base space of $R_{S, Q}$ is efficient by (1-3) above; computing the pre-image configurations of the covering map is efficient by (3) above. This yields a traversal that does not leave the branched covering space, namely the ACR $R_{S, Q}$.

The Cayley configuration methodology   draws upon a rich set of tools from graph rigidity, realization and distance geometry \cite{graver1993combinatorial,sitharam2018handbook}, generalizes to other norms \cite{SitharamWilloughby2015} is closely related to a key finite forbidden minor property called flattenability of graphs \cite{Belk2007,belk2007realizability2,SitharamWilloughby2015}, and leads to directions of independent interest to those areas.
Furthermore, the methodology has been implemented as opensource software (EASAL \cite{Ozkan2018ACMTOMS,Prabhu2020JCIM} and CayMos \cite{Wang2014Caymos,SitharamWang2014Beast}) for respectively molecular and particle assembly modeling and kinematic mechanism analysis and design) and has led to several improvements in those areas, besides efficient algorithms  \cite{Baker2015} for the core problem of solving distance  constraint  systems.

\subsubsection{Cayley Parametrization: Formal Definition and Usage}
We now formally define Cayley parametrization \cite{SitharamGao2010, SitharamWilloughby2015, OzkanBiCoB2011, Ozkan2018ACMTOMS, Prabhu2020JCIM} in the current context. 

Each ACR $R_{S, Q}$ has an underlying \emph{active constraint graph (ACG)} $G_Q= (V_Q, E \cup E_Q)$, whose vertex set $V_Q$ represents the set of the points involved in the  constraints in $Q$. There are two sets of edges:
\begin{itemize}
    \item $E$: all $(a, b)$ where $a$ and $b$ belong to the same point set in $S$.
    \item $E_Q$: pairs $(a, b)\in Q$.
\end{itemize}

Define a \textit{nonedge} of a graph as a vertex pair not in its edge set. One way to represent or parametrize a Cartesian configuration $T$ in an ACR $R_{S,Q}$ is using the tuple of distances/lengths attained in the configuration by a subset $F$ of nonedges of its ACG $G_Q$. This tuple is the corresponding Cayley configuration, with each Cayley coordinate or parameter (value) of the tuple being a nonedge length (value). Formally, this defines an easily computable Cayley parametrization map $\pi_F$ that maps the ACR $R_{S,Q}$ into the Cayley base space $\pi_F(R_{S,Q})$. By choosing a minimal nonedge set $F$ generically of size $(m-\lvert E_Q \rvert)$ whose addition makes the ACG $G_Q$ rigid \cite{sitharam2018handbook}, we ensure that a Cayley configuration corresponds to at most finitely many Cartesian configurations. 

Next, given an ACR $R_{S,Q}$ we describe which ACG’s $G_Q$ have such a minimal, rigidifying set of nonedges $F$, so that the Cayley parametrization maps to a convex Cayley base (configuration) space $\pi_F(R_{S,Q})$. The formal theorem and proof can be found in \cite{SitharamGao2010}, from which a simple algorithm follows for judiciously choosing Cayley parameters $F$. First we define a relevant class of graphs.   

A graph is a \textit{complete 3-tree} if it can be obtained by an inductive construction that starts with a triangle graph and successively adds a new vertex adjacent to the vertices of a triangle in the graph constructed so far. A complete 3-tree has $(3 \lvert V \rvert -6)$ edges and is minimally rigid in $R^3$, therefore it generically has finitely many realizations. 

We find a complete 3-tree $G_Q^*$ that includes all the vertices in $V_Q$, and sets of edges: (1) $E_Q$ (2) a complete 3-tree for each of the vertex sets $V_Q \cap S_i, \forall S_i \in S$. Then the edges in $G_Q^*$ that are not in any of the above 3 sets are chosen as the set $F$ of Cayley parameters. It is shown in \cite{Prabhu2020JCIM, OzkanBiCoB2011} that for almost all ACRs, Cayley parameters satisfying these properties can be found, thereby ensuring a convex Cayley configuration (base) space. Such ACGs $G_Q$ belong to a \textit{nice class} .

In the limiting case where $R_{S, Q}$ is narrow, it is a $(m - \lvert Q \rvert)$-dimensional object in a $m$-dimensional ambient space (Cartesian). Traditional methods of sampling would enforce $Q$ constraints in the ambient $m$-dimensions, thus losing the advantage of the lower intrinsic dimension of $R_{S, Q}$ region and making the sampling computationally complex and challenging, as shown in the 3rd column of Figure 3 of the main article. Further illustrations can be found in \cite{Prabhu2020JCIM} and cover page figure of \cite{Ozkan2021JCTC}.

When Cayley parametrization $\pi_F$ is applied by choosing as Cayley parameters a minimal nonedge set $F$ generically of size $(m-\lvert Q \rvert)$ whose addition makes the ACG $G_Q$ rigid \cite{sitharam2018handbook}, the dimension of $\pi_F(R_{S, Q}) = \lvert F \rvert = m-\lvert Q \rvert$. This is also the intrinsic dimension of a topologically complex, Cartesian $R_{S, Q}$ living in m ambient dimensions. Applying Cayley parametrization essentially “flattens” $R_{S, Q}$ to a simple, convex base Cayley configuration space whose ambient dimension is the same as its intrinsic dimension.

Using these properties, one can efficiently traverse  the Cartesian ACR $R_{S, Q}$ living in high ambient dimension as follows. Traversing the lower dimensional base space of $R_{S, Q}$ is efficient by the 3 nice properties mentioned above; easily compute the pre-image configurations for a given Cayley configuration (lengths for nonedges $F$), 
since for complete 3-trees with  given pair-wise distances for edges, it is  straightforward to find Cartesian configurations satisfying those distance constraints. This yields a traversal that does not leave the branched covering space, namely the ACR $R_{S, Q}$.

Furthermore, the discovery of a child ACR from a parent ACR is easier because of the boundary-interior relationship between these ACRs. When sampling $\pi_F(R_{S, Q})$ using Cayley coordinates, i.e. lengths of nonedges in $F$, when an extra  constraint becomes active   between 2 points $(a, b)$ in different point sets of $S$,   (i.e. at the point when $C_1$ constraint between $a$ and $b$ is violated), binary search can be used to find a configuration close enough to the boundary where the  constraint becomes active, i.e. $dist(a, b)=l(a,b)$, where the inequality in $C_1(a, b)$ becomes an equality. Then the child ACR $R_{S, Q\cup(a, b)}$ is created in the roadmap, and sampling immediately begins on the child, until a leaf or sink ACR of the roadmap DAG is reached. Since the dimension generically drops by 1 from parent to child, for leaf ACR’s, $R_{S, Q}$, $\lvert Q \rvert = m$. Therefore, the \textit{boundaries} of the base space of $R_{S, Q}$ are explicitly detected and traversed. The boundaries represent two types of transitions: (i) the inequalities in $C_1$ and $C_2$ above become tight, or (ii) the real pre-image of the covering map becomes empty (the pre-image is complex); these are additionally the intersections of the branches or flips of the covering space. 

\subsection{Mathematical details of the UC methodology - Configurational Entropy and Cartesian Volume Computation}
\label{sec:algorithm_overviewappendix}
Here we describe three essential parts of UC:

(1) The first part is an algorithm \textit{Uniform Cartesian} that solves Problem 2 when the graph $G_Q$ is in the aforementioned nice class   for $d\le 3$ (as characterized in \cite{SitharamGao2010,Prabhu2020JCIM}) in time linear in the output size, i.e. in the number of $\epsilon$-cubes that intersect $R_{S,Q}$. This is optimal (and nontrivial) since in Cartesian, $R_{S,Q}$ is a topologically complex, effectively lower dimensional subset of the ambient space. In addition to leveraging the known efficient method (*) for sampling the base space of $R_{S,Q}$ in Cayley coordinates \cite{Ozkan2018ACMTOMS,Prabhu2020JCIM}, our algorithm is inspired by a slicing algorithm for 3D printing very large objects filled with mapped (curved) microstructures \cite{YOUNGQUIST2021103102}. See Figures 1, 3, and 4 of the main article. 

(2) The second is of independent interest: a space-efficient grid traversal method that  on average takes sublinear space in the number of grid cubes visited. This indicates sublinear space complexity (in terms of output size) for  modified Problem 1 and 2 that requires \textit{at least} (instead of exactly) one point per grid hypercube that intersects $R_{S,Q}$.

(3) The third is an opensource software implementation of the above algorithm that is used to compare the performance of our method for Problem 1 using variants of the obvious method (*) described above, i.e., Cayley sampling according to 3 different distributions together with pre-image computations of the covering map. These variants were already shown to have significant advantages in efficiency and efficiency-accuracy tradeoffs, in comparison to prevailing Monte Carlo based methods in \cite{Ozkan2021JCTC}. The implementation relies on efficient grid hypercube representations that could be of independent interest: they speed up the extraction of arbitrary dimensional facets and simplices and their intersection with the configuration space $R_{S,Q}$.

 For completeness, we recall the \textbf{input}  to  the algorithm  which includes the  set  $S$ of point sets, the constraints   $C=(C_1, C_2)$ of Section 2.2 of the main article,  ACR $R_{S,Q}$ defined by  the ACG $G_Q$ in the nice class, and
 the required accuracy $\epsilon$ for Problems 1 and 2.
 
 For reasons of exposition and the current software implementation, we make several   {\bf assumptions} that we justify as follows: we assume $d=3$ and $\lvert S \rvert = 2$ whereby the ambient dimension $m=6$.  While smaller $d$ are subsumed, larger $d$ currently lack any useful characterization of the ``nice'' class $\mathcal{C}_d$.   A roadmapping algorithm for $\lvert S \rvert > 2$ is  given in \cite{Prabhu2020JCIM}, and there is no change to the Cayley parametrization and volume computation, but there are more ACGs that fall outside the $\mathcal{C}_d$ and have to be dealt with.    We further assume the modified Problems 1 and 2 that require \textit{at least} (instead of exactly) one point per grid hypercube that intersects $R_{S,Q}$.  From the  solutions  to these modified problems, a straightforward output data structure with a hash map  yields solutions to the original problems.

The algorithm has the 4 following steps.

\begin{enumerate}
\item Using the covering map $\pi_{G_Q}$ (see Section \ref{sec:previous} and \cite{SitharamGao2010},   \textbf{sample the base space} $\pi_{G_Q}(R_{S,Q})$ in Cayley coordinates, via the method in \cite{Ozkan2018ACMTOMS, Prabhu2020JCIM} that determines a Cayley step-size based on $\epsilon$ and finds boundaries and extremal configurations. Further compute the corresponding pre-image Cartesian configurations $s$ in $R_{S,Q}$.  Denote by $s_f$ the pre-image configuration in flip $f$.
\item For each flip $f$, using the Cartesian $\epsilon$-grid hypercube containing $s_f$ as a starting point, \textbf{generate hypercubes $p$ on-demand, and traverse using a key frontier hypercube data structure}.
\item Use the covering map $\pi_{G_Q}$ to generate Cayley cuboid $\pi_{G_Q}(p)$, then \textbf{calculate the intersection with region $\pi_{G_Q}(R_2(S,Q))$ of effective dimension $(m-\lvert Q \rvert$)}; here $R_2(S,Q)$ is the set of configurations satisfying the constraints $Q  \subseteq C_2$, but not necessarily the constraints in $C1$, hence $R_{S,Q} \subseteq R_2(S,Q)$; this generates partly feasible Cayley configurations $c$.
\item Compute the pre-image configurations $\pi_{G_Q, f}^{-1}(c)$, retain only if fully feasible, i.e. only if $C_1$ is satisfied, and \textbf{find and count the corresponding Cartesian grid cube $p'$} if $p'$is in flip $f$. Note that due to linearization error, $p$ may may differ from $p'$. This solves the modified Problems 1 and 2. A straightforward output data structure stores the cubes $p'$ and locates them with a hash map to avoid double counting. This solves the original unmodified problems.
\end{enumerate}
We further describe Steps 2 and 3 in detail in the following paragraphs.

\subsection{Detailed description of hypercube decomposition in Step 3 of UC}
\label{sec:step3appendix}
As described in the main article, step 3 is challenging due to the following reasons:
\begin{itemize}
    \item $\pi_{G_Q}$ is a quadratic map, thus $\pi_{G_Q}(p)$, the mapped cuboid in base space, is a non-linear (but path connected) object, making direct intersection calculation unrealistic;
    \item Intersecting the curved cuboid $\pi_{G_Q}(p)$ and $\pi_{G_Q}(R_2(S,Q))$ yields a potentially disconnected $(m-\lvert Q \rvert)$-dimensional region without a tractable description.
\end{itemize}

To tackle both issues, we provide a series of workaround operations to calculate intersection:
\begin{enumerate}
    \item Instead of using a single, $m$-dimensional $\pi_{G_Q}(p)$ in intersection calculation, decompose $p$ into a collection of $\lvert  \rvert$-dimensional objects $\{p_1, p_2 ... \}$ and map each $p_i$ to $\pi_{G_Q}(p)_i$ in Cayley space. 
    \item Define approximation function $L(\cdot)$ based on linearization, apply it to each $\pi_{G_Q}(p)_i$ to get $L(\pi_{G_Q}(p)_i)$, which is also $\lvert Q \rvert$-dimensional.
    \item Calculate intersection between $L(\pi_{G_Q}(p)_i$ and $\pi_{G_Q}(R_2(S,Q))$ by solving a system of linear equations for a convex combination.
\end{enumerate}
Taking advantage of co-dimensional relationship between $L(\pi_{G_Q}(p)_i)$ ($\lvert Q \rvert$-dimensional) and $\pi_{G_Q}(R_2(S,Q))$ ($(m-\lvert  Q \rvert)$-dimensional), intersections obtained are generically 0-dimensional, i.e. points. After these operations, each $p_i$ generates at most one intersection point $c_i$ due to linearity of both $L(\pi_{G_Q}(p)_i$ and $\pi_{G_Q}(R_2)$. The set of intersection points generated by this process is then input into Step 4 of the overall algorithm.

We hereby provide several different ways of mapping, decomposing, and linearizing $p$ into the set of $L(\pi_{G_Q}(p)_i)$ for varying degrees of accuracy and efficiency in the intersection calculation. 

\subsubsection{Decomposition of Hypercube Based on Simplicial Element}
\label{sec:regulardecomposition}
$\pi_{G_Q}(\cdot)$ maps a Cartesian simplex into a curved simplex in Cayley space. Define linearization $L(\cdot)$ on a curved simplex as rebuilding the simplex with all its vectors; this simplex $\sigma$ will generally preserve its dimensionality after the transformation. To decompose hypercube $p$, the following steps are performed:
\begin{enumerate}
    \item Decompose $p$ into $\lvert  Q \rvert$-dimensional facets.
    \item Decompose each facet $\phi$ into $\lvert  Q \rvert$-dimensional simplices $\sigma$.
    \item Calculate $\pi_{G_Q}(V(\sigma))$, i.e. map the set of vertices of $\sigma$ into Cayley.
    \item Define the Cayley simplex $L(\pi_{G_Q}(\sigma))$ with $\pi_{G_Q}(V(\sigma))$ as vertices.
\end{enumerate}
This definition of $L(\pi_{G_Q}(\cdot))$ on simplices is simple and straightforward while preserving crucial information of simplices. 
While all the variants given here are collectively referred to as {\bf UC} (Uniform Cartesian), the name \textbf{EASAL-UC} is  reserved for the above method  from this point on. This method is illustrated in Figure 5 of the main article. 
However, we observed significant linearization error especially when $p$ is close to the border of $R_{S,Q}$ defined by $C_1$, as well as potential to speed up the process by approximating more aggressively and omitting some of the data. Modifications towards these goals are given below. 

\subsubsection{Modified Decomposition of Simplicial Element - EASAL-face-UC and EASAL-hybrid-UC}
\label{sec:moddecomposition}
Coordinates of $(m-1)$-dimensional face centers are vital in avoiding distortion caused by linearization mentioned in the previous paragraph, see Figure 6 of the main article. We modify the decomposition procedure as follows:
\begin{enumerate}
    \item Decompose $p$ into $(m-1)$-dimensional faces.
    \item Decompose each $(m-1)$-dimensional face  into $(\lvert Q  \rvert -1)$-dimensional facets.
    \item Decompose each facet $\phi$ into $(\lvert Q \rvert-1)$-dimensional simplices $\sigma$.
    \item Define $\lvert Q \rvert$-dimensional Cartesian simplex $\sigma'$ with $V(\sigma)$ and   the corresponding $(m-1)$-dimensional face center.
    \item Calculate $\pi_{G_Q}(V(\sigma'))$.
    \item Build Cayley simplex $L(\pi_{G_Q}(\sigma'))$ using $\pi_{G_Q}(V(\sigma'))$ as vertices.
\end{enumerate}
This method is referred to as \textbf{EASAL-face-UC}. 

The decision of using $(m-1)$-dimensional face centers is to avoid significant distortion due to linearization error  because that decomposition involving $m$-dimensional cube centers does not capture the distortion  due to the Cayley covering map.  See illustration in Figure 6 of the main article. 

This method improves the accuracy at the cost of higher time consumption. To lower overall resource cost, we opt to combine the aforementioned 2 methods. First, regular decomposition is performed on each hypercube, and when it fails to find any feasible intersection, face center version is used. We name this hybrid  combination of  EASAL-UC and face-UC as \textbf{EASAL-hybrid-UC}, and it turns out to be the best performing method overall.

\subsubsection{Decomposition with Facet Parallelepiped Element - EASAL-basis-UC}
\label{sec:basisdecomposition}
 This is a relatively coarse, yet fast variant of our algorithm. To achieve this, we introduce the following method. 

Instead of simplices, we stop decomposition at facet level. Define $L(\phi)$ for Cartesian facet as  a parallelepiped of the same dimension as the facet centered at $\pi_{G_Q}(center(\phi))$ and spanned by a set of basis vectors calculated from $p$. Such an approach retains the property of easy intersection calculation with co-dimensional Cayley region. This method reduces the total number of linear combination calculations to a fraction of previous methods', due to the fact that each facet (instead of simplex) will generate at most 1 intersection.
\begin{enumerate}
    \item From $p$, get all $(m-1)$-dimensional face centers $fc$. 
    \item Map $fc$ into Cayley space to get $\pi_{G_Q}(fc)$ and calculate set of basis vectors using $\pi_{G_Q}(fc_i)$ of opposing faces.
    \item Decompose $p$ into $\lvert Q \rvert$-dimensional facets $\phi$ .
    \item Span the facet image $\pi_{G_Q}(\phi)$ in Cayley space with $\pi_{G_Q}(center(\phi))$ as origin and corresponding basis vectors from Step 2.
\end{enumerate}
We call this variant \textbf{EASAL-basis-UC} as it uses basis vector of parallelepiped to calculate intersection. 

\subsubsection{Towards Loosening the Constraints in $C_2$: $\lvert Q \rvert = 1$ - EASAL-thick-UC}
\label{sec:segmentdecomposition}
 All the variants mentioned above  treat constraints in $C_2$ as    distance  constraints (interval of length 0) and therefore as \textbf{equations}. When we loosen such constraints to be larger intervals, changes have to be made to sample this new ``thick'' $R_{S,Q}$   becomes $m$-dimensional. However,   EASAL provides method to check whether a Cayley point $v$ meets  distance interval constraints. To calculate whether a cube $p$ intersects $\pi_{G_Q}(R_2)$, we map it into a Cayley point set and check if any of the points in the set is within  interval  specified by a constraint in $C_2$. Here we discuss one specific case when $\lvert Q \rvert = 1$, i.e. $C_2$ consists of only 1 distance interval constraint, referred to as 5-dim \textbf{EASAL-thick-UC} in the latter part of this article.
\begin{enumerate}
    \item Decompose $p$ into 1-dimensional facets $\sigma$ following modified decomposition method.
    \item For each $\sigma$, map both endpoints  into Cayley. $L(\pi_{G_Q}(\sigma))$ is a segment in Cayley space.
    \item  Divide $L(\pi_{G_Q}(\sigma))$ equally, to get set of dividing points $v$.
    \item Check each $v$ to see if  $C_2$ constraint is satisfied. 
\end{enumerate}
It is worth noting that expanding the usage of this method to cases where $\lvert Q \rvert > 1$ is  not straightforward, since $L(\pi_{G_Q}(\cdot))$ has no obvious definition. Therefore, only $\lvert Q \rvert = 1$ version is implemented in the software accompanying this paper.

\subsection{Detailed description of Frontier Hypercube Data Structure in Step 2 of UC}
\label{sec:step2appendix}

As mentioned in the main article, the \textit{Frontier Hypercube Graph} data structure is used to store visited hypercubes as well as their relationship with yet-to-visit hypercubes, therefore avoiding repetitions in the traversal. The following is a detailed description for said data structure.
 
First we define a hypercube as \textit{inspected} when one of its $(m-1)$-dimensional face neighbors is processed in Step 3. Visiting of each hypercube is achieved by differentiating between inspected and processed hypercubes and storing only inspected but unprocessed frontier hypercubes in the traversal procedure and discarding interior hypercubes i.e., processed hypercubes whose neighbors have already been inspected. 
Two inspected hypercube containers, $P$ and $NP$, are maintained during procedure, with $P$ for \textit{promising} hypercubes yet to be processed, and $NP$ for \textit{not-yet-promising}. A hypercube is deemed promising if one of its face neighbors has a $\lvert Q \rvert \le m$ dimensional facet containing a valid intersection and pre-image in $R_{S,Q}$ in their shared $(m-1)$ dimensional face, and not-yet-promising if some of its face neighbors have been processed but the hypercube is not in $P$. Each face of a hypercube is given 1 of 3 labels: shared with a processed cube, shared with an uninspected (and thus unprocessed) cube, or shared with inspected but unprocessed cube. $P$ and $NP$ are implemented as a combination of stack and unordered set (hash map) with relative Cartesian coordinates of a hypercube's center as key and a set of labels, one for each face neighbor, as value.
\begin{enumerate}
    \item Initialize $P$ with hypercubes generated in Step 1 and $NP$ empty.
    \item Pick a cube $c$ to process from $P$. In particular, only facets belonging to faces shared with unprocessed cubes are processed to avoid repetition.
    \item Put $c$'s neighbors into $P$ or $NP$, or move them from $NP$ to $P$, based on the result of processing $c$. Face labels of these neighbors are updated. Remove $c$ from $P$.
    \item Algorithm ends when $P$ is empty.
\end{enumerate}
 
To elaborate: it is possible that at the time $c$ is chosen from $P$ to be processed, in fact all of $c$'s faces were shared with previously processed hypercubes, in which case, there is nothing further to be done and $c$ is removed from the data structure. 
It is also possible that although some of $c$'s faces were shared with unprocessed hypercubes in $P$ or $NP$, or uninspected hypercubes, all of $c$'s $\lvert Q \rvert$ dimensional facets could have already been processed. I.e. there is no Step 3 or 4 to be done at the time $c$ is chosen to be processed. 
In any case, faces corresponding to $c$'s uninspected or unprocessed face neighbor hypercubes are processed one face at a time. Effectively any of $c$'s $\lvert Q \rvert$-dimensional facets not shared with processed cubes are processed. The faces corresponding to $c$'s unprocessed neighbors in $P$ or $NP$ are processed and these neighbors' shared faces with $c$ are appropriately relabeled. 
In this process, some hypercubes could move from $NP$ to $P$.
Any of $c$'s uninspected face neighbors that were previously not in the frontier hypercube data structure are now added to $P$ or $NP$, appropriately labeling those shared faces. 
At this point, $c$ is considered processed and is removed from the frontier hypercube data structure. 
 \textit{The key property of this data structure is that a hypercube $c$ can be removed as soon as it is processed without compromising the traversal}.
 Note that $P$ and $NP$ could contain disconnected components and even singleton hypercube/vertices (all of whose face neighbors have either been processed or have not been inspected). Furthermore, shared faces between two hypercubes in $P$ could already contain $\lvert Q \rvert$-dimensional facets that have yielded points in $R_{S,Q}$: this is because such facets could additionally belong to faces shared with already processed cubes. However, faces corresponding to edges incident on any hypercube in $NP$ cannot contain such a facet.

\subsection{Complexity Analysis}
\label{sec:complexityappendix}

Here we provide a time and space complexity analysis which is the same for all UC variants.  
 Since  we deal with a sampling algorithm,  it is appropriate that our complexity bounds are given in the  number of (essential) output samples,  rather than the size of the input.  It should be noted that since unused or unnecessary samples are not part of the output, the time complexity gets worse with the number of  unnecessary samples utilized by any given algorithm.  All the UC variants attempt to minimize the number of unnecessary samples.

\noindent
\textbf{Theorem:} Let $m=6(k-1)$ be the ambient dimension (degrees of freedom) of configuration space for a system with $k$ rigid bodies. Let the volume of an ACR $R_{S,Q}$ at energy level (effective dimension)  $D<m$ be $l^D$, where $l$ is some one-dimensional length measure, and the number of (essential) output samples is $O(l^D)$. Then:
\begin{itemize}
    \item Using Cayley parametrization lowers the time cost from $O(l^m)$ to $O(l^D)$, i.e. linear  time complexity in the number of output samples;
    \item Adopting the frontier hypercube data structure lowers the average space cost of UC from $O(l^D)$ to $O(l^{D-1})$, i.e., sublinear  space complexity in the number of output samples.
\end{itemize}

\noindent
\textbf{Proof:} $R_{S,Q}$ resides in $m$-dimensional Cartesian  ambient space and have   arbitrarily complex topology therefore sampling it via  traditional  methods in $m$-dimensional  ambient space  could in the worst case result in a  time complexity  of $O(l^m)$. Mapping it to an effectively $D$-dimensional Cayley base space and sampling it with a non-repetitive, stop-at-boundary traversal has the same asymptotic complexity as its volume, i.e. $O(l^D)$. For space complexity, brute-force traversal and notetaking (to avoid repetition leading to infinite loop) will record the entirety of $R_{S,Q}$ eventually when the process finishes, thus taking $O(l^D)$ space. The frontier hypercube data structure only keeps track of the surface of the visited part of $R_{S,Q}$, which, for an average $D$-dimensional region $R_{S,Q}$ costs $O(l^{D-1})$-dimensional memory to store.

In contrast,  some methods using Monte Carlo sampling  take $>O(l^m)$ time, including many discarded and repeated samples, but only $O(1)$ space, since they do not ``remember'' any visited regions.  

\begin{center}
    \begin{tabular}{|c|c|c|}
\hline
     & UC & Monte Carlo \\
     \hline
    Time complexity & $ O(l^D)(linear\  in\  output\  size) $ & $ \ge O(l^{m>>D}) $ \\
    \hline
    Space complexity & $ O(l^{D-1})(sublinear\ in\ output\ size) $ & $ O(1) $ \\
    \hline
\end{tabular}
\end{center}

  \bibliographystyle{elsarticle-num} 
  \bibliography{bibliography}





\end{document}